\documentclass[aps,prc,reprint,superscriptaddress,nofootinbib]{revtex4-2}

\usepackage{graphicx}
\usepackage{dcolumn}
\usepackage{amsmath}
\usepackage{amssymb}
\usepackage{bm}
\usepackage{color}
\usepackage[normalem]{ulem}
\usepackage[colorlinks=true, linkcolor=blue, citecolor=blue, urlcolor=magenta]{hyperref}
\usepackage{xcolor}

\newcommand{\Wsur}{W_{\rm sur}}
\newcommand{\Ecm}{E_{\rm c.m.}}
\newcommand{\Ecn}{E^{*}_{\rm CN}}
\providecommand{\lambdabar}{\ensuremath{\mathchar'26\mkern-9mu\lambda}}

\begin{document}

\title{The effect of the coupling between the neutron multiplicity and the
fission competition on the evaporation residue cross sections of the
$^{48}$Ca\,$+\,^{244}$Pu reaction}

\author{Bakhodir Kayumov}
\email[Corresponding author: ]{b.kayumov@newuu.uz}
\affiliation{Department of Physics, New Uzbekistan University, Tashkent 100000, Uzbekistan}
\affiliation{Laboratory of Theoretical Nuclear Physics, Institute of Nuclear Physics, Tashkent 100214, Uzbekistan}

\author{Sarvar Amirov}
\affiliation{Department of Physics, New Uzbekistan University, Tashkent 100000, Uzbekistan}

\begin{abstract}
The survival probability $W_{\rm sur}$ of an excited compound nucleus
against fission is a key factor determining the evaporation-residue
cross section in the synthesis of superheavy nuclei. The 
$W_{\rm sur}$ for $^{292}$Fl formed in the
$^{48}$Ca+$^{244}$Pu reaction was  investigated, with particular attention to the
coupling between neutron multiplicity and fission competition during
the de-excitation cascade. The entrance-channel capture and fusion
cross sections are obtained within the dinuclear system model, while
the neutron--fission competition is treated using the
Vandenbosch--Huizenga formalism with the Kramers and Strutinsky
corrections and a fission barrier damped with temperature and angular
momentum. The survival probability is calculated using two treatments
of the evaporation--fission cascade. In the conventional factorized
approach, the realization probability of a given $xn$ channel is
treated independently of the neutron--fission competition, with the
branching ratios evaluated along a mean excitation-energy trajectory.
In the coupled approach, the full excitation-energy population is
propagated through the cascade, so that neutron multiplicity and
competition with fission are determined simultaneously at the energy
actually reached by each nucleus. The comparison shows that the
factorized treatment underestimates the $3n$ survival probability by
factors of 17 and 24 at $E^*=39$ and 47 MeV, respectively, while its
effect on the $4n$ channel is considerably smaller. Thus, the coupling
primarily affects the relative strengths and shapes of neighboring
evaporation channels rather than their common normalization. The
resulting $3n$ and $4n$ evaporation-residue excitation functions
reproduce eight of the fourteen measured cross sections within a
factor of two. A comparison of two prescriptions for the cold fission
barrier further shows that differences of less than 1 MeV can change
$W_{\rm sur}$ by more than an order of magnitude. These results
demonstrate the importance of treating neutron multiplicity and
fission competition consistently in calculations of superheavy-nucleus
survival and identify the fission barrier as a major source of
uncertainty in quantitative predictions of evaporation-residue cross
sections.
\end{abstract}


\maketitle

\section{Introduction}

The synthesis of superheavy elements (SHEs) is one of the important directions
of modern nuclear physics, since it gives information on the stability and
structure of nuclei at the limits of the periodic table
\cite{Hofman2000,Hofmann2016}. The heaviest elements have been obtained in the
hot fusion reactions with a $^{48}$Ca beam
\cite{Oganessian2007,Oganessian2013,Oganessian2023}, and the isotopes of
flerovium, which are formed in the $^{48}$Ca\,$+\,^{244}$Pu reaction, are among
the most thoroughly measured of them
\cite{Oganessian2004b,Oganessian2004a,Dullmann2010,Gates2011,SamarkRoth2023,Gates2026}.
Since the targets heavier than $^{249}$Cf are not available, the further
extension of the periodic table requires projectiles heavier than $^{48}$Ca, and
the first result of this kind has been reported recently for the production of
livermorium with a $^{50}$Ti beam \cite{Gates2024}. The cross sections of such
reactions are of the order of a picobarn or below
\cite{Khuyagbaatar2020,Tanaka2022}, so that the reliability of the theoretical
estimates of each factor entering them becomes essential.

It is well known that the ER cross section is presented as a product of three
factors: the capture cross section of the projectile by the target nucleus, the
probability $P_{\rm CN}$ of the CN formation in competition with quasifission,
and the survival probability $\Wsur$ of the excited CN against fission during
the emission of $x$ neutrons. The first two factors are determined by the
entrance channel and are calculated in this work within the dinuclear system
(DNS) model \cite{Volkov1978,Antonenko1995,Nasirov2005}, which has been applied
to many reactions leading to heavy and superheavy nuclei
\cite{Nasirov2019,Kayumov2022,Kayumov2025,Nasirov2026NPA}. The
same three-factor decomposition is used in the other theoretical approaches
developed for the SHE synthesis
\cite{Zagrebaev2008,Liu2013,Bao2015,Li2022120,Zhu2023,Zhang2024}, and the
predictions obtained with them for the same reaction may differ by one or two
orders of magnitude. The third factor is determined by the de-excitation of the
already formed CN and is the subject of the present work.

It should be stressed that $\Wsur$ may be as small as $10^{-10}$ for the nuclei
with $Z \geq 114$, because the fission barrier of a superheavy nucleus is
entirely of shell origin and is comparable with the neutron separation energy.
The ratio $B_f/B_n$ along the cascade, and the rate at which the shell
contribution to $B_f$ is washed out with the temperature and the angular
momentum, therefore control the result. It is important to note that the small
value of $\Wsur$ is the product of several branching ratios, each of which is
close to $10^{-2}$ or below, and that an error of a few hundred keV in the
barrier of any nucleus of the chain is amplified exponentially by the
Bohr--Wheeler width.

The neutron--fission competition along the cascade is treated here by the
Vandenbosch--Huizenga expression \cite{Vandenbosch1973}, which is standard in
the statistical model. Together with a DNS entrance channel the survival
probability is evaluated either by propagating the population of the cascade
over the excitation energy, or by the analytic expression in which it is
factorized into the probability of realizing the $xn$ channel and a product of
branching ratios taken along a mean trajectory of the cascade, so that the
multiplicity of the emitted neutrons and the competition with fission are
treated as independent \cite{Nasirov2019,Kayumov2022}. It is this second form
which is examined here. The factorization is not exact,
since fission removes the population preferentially
where the excitation energy is high, and the nuclei which survive to emit a
further neutron are colder than the mean trajectory indicates. In the present
work the factorization is removed and the whole population of the cascade is
propagated over the excitation energy.

The main goal of this work is therefore to present, in a form which can be
reproduced, the statistical treatment of $\Wsur$ used together with the DNS
calculations, and to establish which of its ingredients the result is sensitive
to. Particular attention is paid to the fission barrier, for which two
independent prescriptions are compared, since it is the quantity that enters the
neutron--fission competition exponentially. All the nucleus-dependent quantities
are taken from the FRDM(2012) tables \cite{Moller2016,Moller2015}. The only
quantity of the de-excitation model adjusted in this work is the damping
temperature $T_0$ of Eq.~(\ref{eq:damping}); the constant $K_0$ of
Eq.~(\ref{eq:gnf}) and the Kramers and Strutinsky factors are kept at the values
adopted in the works from which they are taken, and the excitation-energy scale
of the DNS calculation is used without adjustment. The method is demonstrated
for $^{292}$Fl formed in the $^{48}$Ca\,$+\,^{244}$Pu reaction, for which the
$3n$ and $4n$ channels have been measured with four separators at three
laboratories over an interval of about $20$~MeV of the excitation energy.

This paper is structured as follows: Sec.~\ref{sec:method} describes the
entrance-channel input taken from the DNS model, the statistical treatment of
the de-excitation, the level-density parameter, the two fission-barrier
prescriptions and the coupled treatment of the cascade, and
Sec.~\ref{sec:results} discusses the results obtained for the
$^{48}$Ca\,$+\,^{244}$Pu reaction. Finally, the conclusions are reported in
Sec.~\ref{sec:conclusion}.

\section{The method}\label{sec:method}

\subsection{Entrance-channel input from the dinuclear system model}

In the DNS approach the synthesis of a superheavy nucleus is treated as three
consecutive and separable stages: capture of the projectile by the target with
full momentum transfer, evolution of the captured dinuclear configuration
towards a compound nucleus in competition with its decay into two fragments,
and de-excitation of the heated and rotating CN
\cite{Antonenko1995,Giardina2000,Kayumov2022}. The first two stages provide the
input of the present work and are summarized here.

The capture is not described by a barrier-penetration formula alone. The number
of partial waves leading to capture has been calculated by solving the classical
dynamical equations for the relative distance $R(t)$ between the centres of mass
of the interacting nuclei, together with the equations for the angular momentum
of the relative motion and of the intrinsic rotation of the fragments
\cite{Giardina2000,Nasirov2005}. The radial equation contains a friction term
$\gamma_R \dot{R}$ and the conservative force $-\partial V/\partial R$, while the
tangential friction $\gamma_\theta$ transfers orbital into intrinsic angular
momentum. The friction coefficients are not free parameters but are calculated
microscopically from the particle--hole excitations and the nucleon-exchange rate
between the two nuclei \cite{Adamian1997,Nasirov2005}, and the nucleus--nucleus
potential is obtained by the double-folding procedure with the effective
density-dependent nucleon--nucleon forces \cite{Migdal1983,Nasirov2005}.

Capture is registered when enough of the initial kinetic energy has been
dissipated for the trajectory to be trapped in the pocket of $V(R)$, so that the
system can no longer re-cross the Coulomb barrier $V_B$ from inside; otherwise a
deep-inelastic collision takes place and no dinuclear system is formed. This
condition defines a dynamical limiting angular momentum $\ell_d$. For the
reaction considered here the capture cross section is non-zero from
$\Ecm \simeq 180$~MeV upwards, and the partial waves which contribute to
capture extend to $100\hbar$ at the highest energies considered. For
$V_{\rm min}(\ell) < \Ecm < V_B(\ell)$ capture proceeds by tunnelling, the
penetrability being evaluated in the WKB approximation \cite{Kemble1935}. The
depth of the pocket, $B_{\rm qf}=V_B-V_{\rm min}$, is identified with the
quasifission barrier and measures the stability of the dinuclear system against
re-separation \cite{Nasirov2005,Kayumov2022}. Because $^{244}$Pu is strongly
deformed, $V_B$, $V_{\rm min}$ and therefore $\ell_d$ depend markedly on the
orientation angle $\alpha_T$ of its symmetry axis relative to the beam, so that
all partial cross sections are calculated at fixed $\alpha_T$ and afterwards
averaged over $\sin\alpha_T\,d\alpha_T$ \cite{Nasirov2005}, the surface
vibrations of the spherical $^{48}$Ca being included as in
Ref.~\cite{Nasirov2024}.

The second stage is described as a diffusion in the charge and mass asymmetry
over the potential energy surface, the valley of which defines the driving
potential \cite{Fazio2003,Kayumov2022}. The charge distribution $D_Z$ of the
dinuclear system is obtained by solving the transport master equation whose
nucleon transition coefficients follow from the occupation numbers and
single-particle energies of the two fragments \cite{Kayumov2022,Fazio2005}. For
each charge asymmetry the system may move towards $Z=0$ over the intrinsic
fusion barrier $B^{*}_{\rm fus}$ and form the CN \cite{Antonenko1995,Adamian2000},
decay over the quasifission barrier $B_{\rm qf}$, or drift towards symmetry over
$B_{\rm sym}$. The branching between the three is evaluated as a ratio of level
densities, which for a Fermi gas reduces to a ratio of Boltzmann factors
$\exp(-B/T_Z)$ with the temperature $T_Z$ of the configuration of charge
asymmetry $Z$, and the fusion probability follows as
$P_{\rm CN}=\sum_Z D_Z P_{\rm CN}^{(Z)}$ \cite{Nasirov2019}. A detailed
description of the calculation of $P_{\rm CN}$ in this approach, with the same
treatment of the driving potential and of the master equation as is used here,
is given in Ref.~\cite{Kayumov2025}. It should be noted
that $P_{\rm CN}$ is not directly observable, and that the values extracted from
the data depend on the assumptions made in separating quasifission from fusion
fission \cite{Giardina2018,Loveland2015}.

The partial capture cross section is obtained from the partial waves which
satisfy the capture condition \cite{Kayumov2025},
\begin{equation}
\begin{aligned}
\sigma_{\rm cap}(\Ecm,\ell) &= \pi \lambdabar^{2}\,(2\ell+1)\,
P_{\rm cap}(\Ecm,\ell), \\
\lambdabar^{2} &= \frac{\hbar^{2}}{2\mu \Ecm},
\end{aligned}
\label{eq:scap}
\end{equation}
where $\mu$ is the reduced mass of the entrance channel. Above the barrier
$P_{\rm cap}=1$ for $\ell<\ell_d$, in the sub-barrier region $P_{\rm cap}$ is
the penetrability, and outside this window it vanishes. The partial fusion
cross section is then
\begin{equation}
\sigma_{\rm fus}(\Ecm,\ell) = \sigma_{\rm cap}(\Ecm,\ell)\,
P_{\rm CN}(\Ecm,\ell),
\label{eq:sfus}
\end{equation}
and the fusion cross section itself is its sum over the partial waves. It should
be noted that $\sigma_{\rm fus}(\Ecm,\ell)$ of Eq.~(\ref{eq:sfus}), which is the
quantity tabulated by the DNS code and used below in Eq.~(\ref{eq:sigma}),
already carries the factor $\pi\lambdabar^{2}(2\ell+1)$. No further statistical
weight is therefore applied in the partial-wave sum, apart from the step
$\Delta\ell$ of the grid on which the DNS results are tabulated. The part of
$\sigma_{\rm fus}$ with $\ell > \ell_f$, where $\ell_f$ is the angular momentum
at which the fission barrier of the CN vanishes \cite{Sierk1986}, is assigned to
fast fission and does not lead to evaporation residues
\cite{Kayumov2022,Graeger2010}. This cut is already applied in the DNS output
used here, $\sigma_{\rm fus}$ vanishing above $\ell_f$, and the partial waves
which remain extend to $40\hbar$ of entrance-channel angular momentum.

Two angular momenta have to be distinguished, both of which are tabulated by the
DNS code. The first is the entrance-channel $\ell$, which
labels the partial waves and fixes the step $\Delta\ell$ of the grid in
Eq.~(\ref{eq:sigma}). The second is the angular momentum $\ell_{\rm CN}$ of the collective
rotation which remains after a part of the orbital angular momentum of the
relative motion has been transferred to the intrinsic rotation of the two
fragments by the tangential friction. In the sticking limit the two are
related by the ratio of the moments of inertia, which gives
$\ell_{\rm CN}=\tfrac{5}{7}\,\ell$ for the dinuclear configuration tabulated
here, so that the surviving partial waves correspond to $\ell_{\rm CN}$ from
$0$ to $29\hbar$. It is this second quantity which enters $\Wsur$, both through
the damping $q(\ell)$ of Eq.~(\ref{eq:damping}) and through the rotational
energy which is removed from the excitation energy. The excitation energy of
the compound nucleus is defined for each partial wave as
\begin{equation}
\begin{aligned}
\Ecn(\Ecm,\ell) &= \Ecm + Q_{gg} - E_{\rm rot}(\ell_{\rm CN}), \\
E_{\rm rot}(\ell_{\rm CN}) &= \frac{\hbar^2 \ell_{\rm CN}^{2}}{2\Im},
\end{aligned}
\label{eq:ecn}
\end{equation}
where $\Im$ is the rigid-body moment of inertia of a sphere of mass number
$A=292$, for which $\hbar^2/2\Im = 2.91$~keV. The rotational energy is
subtracted because it is not available as thermal excitation for the particle
emission. Both the rotational energy and the damping $q(\ell)$ are therefore
evaluated with the same angular momentum, that of the compound nucleus.

The excitation energy is computed with the ground-state reaction $Q$ value of
the DNS calculation, $Q_{gg}=-161.32$~MeV. The FRDM(2012) mass excesses
\cite{Moller2016} give $-160.52$~MeV
and the earlier tables \cite{AUDI2003,Moller1995} a similar value, the mass of
$^{292}$Fl not having been measured. The difference of $0.8$~MeV is of the size
expected from the mass models themselves \cite{Kawai2026}, and it sets the
accuracy within which the excitation-energy scale of such a calculation is
known.

It is necessary to comment on the averaging over the orientations. For a
deformed target the partial cross sections are computed at fixed $\alpha_T$ on a
grid of seven angles, $0^{\circ}$ to $90^{\circ}$ in steps of $15^{\circ}$, and
are averaged
over $\sin\alpha_T\,d\alpha_T$. The seven computed angles are the nodes of that
grid and define six intervals, which are integrated by the trapezoidal rule
with the weights $\cos\alpha_{k-1}-\cos\alpha_k$, whose sum is unity. It is
important to note that this measure is far from uniform in $\alpha_T$: the
tip-on orientations, which have the lowest Coulomb barrier and therefore
dominate the fusion cross section below the barrier, occupy only a small solid
angle. The sub-barrier slope of $\sigma_{\rm fus}$, and through it the width of
the calculated excitation functions, is consequently sensitive to the weighting
with which the orientations are averaged.

The cross section of the $xn$ evaporation channel is finally obtained as
\begin{equation}
\begin{split}
\sigma_{xn}(\Ecm) = \sum_{\ell} \Delta\ell \;
&\sigma_{\rm fus}(\Ecm,\ell) \\
&\times \Wsur\bigl(\Ecn(\Ecm,\ell),\,x,\,\ell\bigr),
\end{split}
\label{eq:sigma}
\end{equation}
where $\Delta\ell$ is the step of the partial-wave grid of the DNS output. It is
important to note that $\Wsur$ is evaluated at the $\ell$-dependent excitation
energy of Eq.~(\ref{eq:ecn}), and not at $\Ecm+Q_{gg}$. The angular momentum is
kept constant along the cascade, since each emitted neutron removes on average
about $1\hbar$, which is small in comparison with the width of the partial-wave
distribution.

\subsection{Survival probability}

The survival probability is the product of the realization probability $P_{xn}$
of the $xn$ channel and of the branching ratios of the neutron emission over
fission at every step of the evaporation cascade,
\begin{equation}
\Wsur(E^{*},x,\ell) = P_{xn}(E^{*},x) \prod_{i=1}^{x}
\left[\frac{\Gamma_n}{\Gamma_n+\Gamma_f}\right]_{i},
\label{eq:wsur}
\end{equation}
where $\Gamma_n$ and $\Gamma_f$ are the neutron emission and fission widths of
the $i$-th nucleus of the cascade. The first factor is calculated by the Jackson
formula \cite{Jackson1956},
\begin{equation}
P_{xn} = I(\Delta_x,\,2x-2) - I(\Delta_{x+1},\,2x),
\label{eq:pxn}
\end{equation}
where $I(z,n)$ is Pearson's form of the regularized incomplete gamma function,
$I(z,n)=\gamma(n+1,z)/\Gamma(n+1)$, so that the second argument is one less than
the shape parameter; for $x=1$ the first term is to be read as unity whenever
$\Delta_1>0$ and as zero otherwise,
\begin{equation}
\Delta_x = \frac{E^{*}-x\bar{B}_n}{T}, \qquad
\Delta_{x+1} = \frac{E^{*}-x\bar{B}_n-B_{x+1}}{T},
\label{eq:jacksondelta}
\end{equation}
$\bar{B}_n$ being the neutron separation
energy averaged over the $x$ nuclei of the cascade and $B_{x+1}$ that of the
$(x+1)$-th neutron. The temperature is evaluated at the middle of the cascade,
that is for the nucleus $A_0-\lfloor x/2\rfloor$ at the excitation energy
$E^{*}-\tfrac{1}{2}x\bar{B}_n$. It is this second term which
removes the events that go on to emit a further neutron, and which makes the
series telescope. The sum $\sum_x P_{xn}$ is then unity to within a few per
cent, the residual departure coming from the fact that the temperature of
Eq.~(\ref{eq:jacksondelta}) is evaluated separately for each channel.

A refinement of the last step has been proposed in Ref.~\cite{Rahmatinejad2023},
in which the fission barrier of the residue replaces $B_{x+1}$ in the second
term for those residues which can fission but can no longer emit a neutron, that
is when $B_f<B_n$. The substitution is conditional, and for the present chain
the condition is never met: $B_f>B_n$ for every nucleus of this cascade at all
relevant residual energies, so that a residue hot enough to fission is also hot
enough to emit an $(x+1)$-th neutron and is already removed by the second term
of Eq.~(\ref{eq:pxn}) as it stands. Applying the substitution unconditionally,
which is not what is proposed there, would leave the $(x+1)n$ events inside the
$xn$ channel and raise $\sum_x P_{xn}$ to as much as $1.9$. The prescription of
Ref.~\cite{Rahmatinejad2023} is therefore inactive here rather than rejected.
The fission of the emitting nuclei is accounted for by the product in
Eq.~(\ref{eq:wsur}).

The competition between the neutron evaporation and the fission at each step is
treated by the Vandenbosch--Huizenga expression \cite{Vandenbosch1973}, which
follows from the Weisskopf \cite{Weisskopf1937} and the Bohr--Wheeler
\cite{Wheeler1939} widths,
\begin{equation}
\begin{split}
\frac{\Gamma_n}{\Gamma_f} =
\frac{4A^{2/3}a_f\,(E^{*}-B_n)}
{K_0\,a_n\left[2\sqrt{a_f(E^{*}-B_f)}-1\right]}
\;\frac{1}{K\,S} \\
\times\;
e^{\,2\sqrt{a_n(E^{*}-B_n)}\,-\,2\sqrt{a_f(E^{*}-B_f)}},
\end{split}
\label{eq:gnf}
\end{equation}
where $A$ is the mass number of the emitting nucleus, while $a_n$ and $a_f$ are
the level-density parameters of the residual nucleus after the neutron emission
and of the fissioning nucleus at the saddle point. The denominator of
Eq.~(\ref{eq:gnf}) vanishes at $E^{*}-B_f=1/(4a_f)$ and is negative below it,
where the asymptotic form used here is in any case not valid. In the
implementation the fission channel is closed and the branching ratio set to unity
whenever $2[a_f(E^{*}-B_f)]^{1/2}\le1$, which keeps the probabilities
non-negative; the region concerned lies within a few keV of the barrier and
carries no weight in the integrated yields.
The fission width carries two further corrections \cite{Kawai2026}.
The first is the Kramers factor \cite{Kramers1940}, which accounts for the
dynamical reduction of the fission width by the friction of the collective
motion,
\begin{equation}
K = \sqrt{1+\left(\frac{\gamma}{2\omega_{\rm sd}}\right)^{2}}
    - \frac{\gamma}{2\omega_{\rm sd}},
\label{eq:kramers}
\end{equation}
and the second is the correction proposed by Strutinsky \cite{Strutinsky1973}
to the Bohr--Wheeler width, which accounts for the different number of
stationary collective states at the ground state and at the saddle point,
\begin{equation}
S = \frac{\hbar\omega_{\rm g.s.}}{T},
\qquad T = \sqrt{\frac{E^{*}}{a_n}} .
\label{eq:strutinsky}
\end{equation}
Both are used with the values adopted in Ref.~\cite{Kawai2026},
$\gamma/2\omega_{\rm sd}=1$ and $\hbar\omega_{\rm g.s.}=1$~MeV, so that
$K=\sqrt{2}-1=0.414$, and neither is adjusted here. Their product raises
$\Gamma_n/\Gamma_f$ by a factor $2.41\,T/\hbar\omega_{\rm g.s.}$, which grows
with the excitation energy and therefore acts on the slope of the excitation
function and not merely on its normalization. For $E^{*} \leq B_f$ the fission
width is negligible and the branching ratio is set to unity.

The remaining ingredients of Eq.~(\ref{eq:gnf}) are not adjusted either. No
empirical enhancement is applied to $a_f$, which is taken as the asymptotic
value of Eq.~(\ref{eq:atilde}) evaluated at the saddle-point deformation. The
resulting ratio $a_f/a_n$ is nevertheless not unity, since $a_n$ carries the
Ignatyuk shell damping and $a_f$ does not, and for the nuclei of this cascade it
lies between $1.27$ and $1.34$ at $E^{*}=35$~MeV. The constant
$K_0 = \hbar^{2}/2m_n r_0^{2}$ equals $14.39$~MeV for $r_0=1.2$~fm, which is the
value usually quoted with Eq.~(\ref{eq:gnf}) and the one used here; it is not
treated as an adjustable quantity, although the ER cross section is proportional
to $1/K_0$ at every step of the cascade. The damping temperature $T_0$ of
Eq.~(\ref{eq:damping}) is therefore the only parameter of the de-excitation
model which has been fitted to the data.

Equation~(\ref{eq:wsur}) is written here in its usual factorized form, in which
the multiplicity of the cascade and the neutron--fission competition are treated
as independent. That factorization is not exact, and the consequences of
removing it are the subject of Sec.~\ref{sec:coupled}.

\subsection{Level-density parameter}

The asymptotic level-density parameter is taken from the four-term expression of
Nerlo-Pomorska \textit{et al.} \cite{Pomorska2006},
\begin{equation}
\begin{split}
\tilde{a} = 0.092\,A &+ 0.036\,A^{2/3}B_s + 0.275\,A^{1/3}B_k \\
                     &- 0.00146\,\frac{Z^2}{A^{1/3}}B_c ,
\end{split}
\label{eq:atilde}
\end{equation}
in which the surface, the curvature and the Coulomb terms are corrected by the
deformation-dependent factors \cite{Hasse}
\begin{align}
B_s &= 1+\tfrac{2}{5}\alpha_2^2-\tfrac{4}{105}\alpha_2^3-\tfrac{66}{175}\alpha_2^4,
\nonumber\\
B_k &= 1+\tfrac{2}{5}\alpha_2^2+\tfrac{16}{105}\alpha_2^3-\tfrac{82}{175}\alpha_2^4,
\label{eq:bsbkbc}\\
B_c &= 1-\tfrac{1}{5}\alpha_2^2-\tfrac{4}{105}\alpha_2^3+\tfrac{51}{245}\alpha_2^4,
\nonumber
\end{align}
with $\alpha_2 = \sqrt{5/4\pi}\,\beta_2$. For $a_n$ these factors are evaluated
at the ground-state deformation $\beta_2$ of the residual nucleus
\cite{Moller2016}, and for $a_f$ at the saddle-point deformation obtained from
the fissility parameter $x_f$ \cite{Hasse},
\begin{equation}
\begin{aligned}
\alpha_2^{\rm sad} &= \tfrac{7}{3}y-\tfrac{938}{765}y^{2}
                      +9.4998\,y^{3}-8.0509\,y^{4}, \\
y &= 1-x_{f},
\end{aligned}
\label{eq:saddle}
\end{equation}
where $x_{f} = Z^2/\left[49.22\,A\,(1-0.3803\,I^{2}-20.489\,I^{4})\right]$ and
$I=(A-2Z)/A$, the ground-state value being added whenever
Eq.~(\ref{eq:saddle}) returns a smaller one. It should be noted that no shell
correction is applied to $a_f$. The saddle-point shape of Eq.~(\ref{eq:saddle})
is only weakly deformed for these nuclei, $\alpha_2^{\rm sad}=0.074$ for
$^{292}$Fl, but it is not the ground-state shape, and the ground-state shell
structure which produces $\delta W$ is not carried over to it. The deformation
entering $a_f$ is therefore that of Eq.~(\ref{eq:saddle}) and is distinct from
the saddle configuration underlying the tabulated barrier of
Ref.~\cite{Moller2015}, which is obtained from a multidimensional
minimization.

The energy dependence of the level-density parameter of the residual nucleus is
included by the Ignatyuk prescription \cite{Ignatyuk75},
\begin{equation}
\begin{aligned}
a(E^{*}) &= \tilde{a}\left[1+\frac{\delta W}{E^{*}}
\left(1-e^{-E^{*}/E_d}\right)\right], \\
E_d &= 20~{\rm MeV},
\end{aligned}
\label{eq:ignatyuk}
\end{equation}
where the shell correction $\delta W$ is taken for every nucleus of the cascade
from the FRDM(2012) table \cite{Moller2016}. The nuclear temperature is
$T=\sqrt{E^{*}/a(E^{*})}$ throughout.

\subsection{Fission barrier}

The fission barrier of a superheavy nucleus is due entirely to the shell
correction, and this contribution melts as the nucleus is heated and as it is
set into rotation. Following the prescription used in our previous work
\cite{Giardina2018,Mandaglio2018,Kayumov2022}, in which the shell contribution
is damped in the temperature and in the angular momentum by two Fermi
functions, the barrier is written as
\begin{equation}
B_f(A,Z,E^{*},\ell) = B_f^{(0)} + \delta W\left[1-h(T)\,q(\ell)\right],
\label{eq:bf}
\end{equation}
with the two damping functions
\begin{equation}
\begin{aligned}
h(T) &= \left\{1+e^{(T-T_0)/d}\right\}^{-1}, \\
q(\ell) &= \left\{1+e^{(\ell-\ell_{1/2})/\Delta\ell_d}\right\}^{-1},
\end{aligned}
\label{eq:damping}
\end{equation}
where $B_f^{(0)}$ is the cold fission barrier of M\"oller \textit{et al.}
\cite{Moller2015}, which already contains the shell contribution, and
$T_0 = 1.20$~MeV, $d = 0.30$~MeV, $\ell_{1/2}=20\hbar$ and
$\Delta\ell_d=3\hbar$. Since $\delta W < 0$ for the Fl isotopes, the second term
of Eq.~(\ref{eq:bf}) progressively removes the shell contribution as the
temperature and the angular momentum increase, leaving the small macroscopic
barrier.

Two prescriptions for $B_f^{(0)}$ are compared in this work. The first is the
tabulated microscopic--macroscopic barrier of M\"oller \textit{et al.}
\cite{Moller2015}, obtained from a full potential-energy surface in a
multidimensional deformation space. The second follows the topographical theorem
of Myers and \'Swi\c{a}tecki, in which the barrier is written as the sum of the
macroscopic barrier and the ground-state microscopic correction,
\begin{equation}
B_f^{(0)} = B_{\rm mac}(Z,A) - \delta W ,
\label{eq:topo}
\end{equation}
with $B_{\rm mac}$ taken from the analytic fit to the Lublin--Strasbourg drop of
Ivanyuk and Pomorski \cite{Lublin}. Since $\delta W<0$, the microscopic term
raises the barrier. For $Z=114$ the macroscopic barrier has already vanished,
$B_{\rm mac} \simeq 0$, so that Eq.~(\ref{eq:topo}) reduces to
$B_f^{(0)} \simeq |\delta W|$ and the barrier is of purely shell origin. It
should be noted that the fit of Ref.~\cite{Lublin} was adjusted for
$35 \leq Z \leq 105$, so that its use at $Z=114$ is an extrapolation; the
polynomial in fact returns $B_{\rm mac}=-0.001$~MeV there, which is negligible
and is retained as it stands.

Since the same damping term is applied in both cases, the two prescriptions
differ by a constant offset at every $E^{*}$ and every $\ell$. Their limits at
high temperature are, however, not equivalent: with Eq.~(\ref{eq:topo}) the
barrier tends to $B_{\rm mac}\simeq0$, which is the expected completely washed
out limit, whereas with the tabulated barrier it tends to
$B_f^{(0)}+\delta W = 0.53$~MeV for $^{292}$Fl, a residue which reflects the
fact that the tabulated saddle energy is not exactly $B_{\rm mac}-\delta W$.

The damping temperature is the least certain input of the model, and the value
$T_0 = 1.20$~MeV used here has been chosen so that the calculated cross
sections agree in magnitude with the measured ones. It should be stressed how
strong this dependence is. The same prescription has been used with
$T_0 = 1.16$~MeV elsewhere \cite{Giardina2018,Mandaglio2018}; at $E^{*}=40$~MeV
that choice lowers $B_f$ of $^{292}$Fl by $0.29$~MeV, which is smaller than the
difference between the two barrier prescriptions discussed above, and it lowers
$\sigma_{xn}$ near the maximum by a factor of $2.5$ for the $3n$ channel and of
$2.9$ for the $4n$ channel. Together with $K_0$, $T_0$ is
therefore the quantity which fixes the absolute scale of the calculation.

The rotational lowering of the barrier is not added separately, since the
rotational energy is already subtracted in $\Ecn(\Ecm,\ell)$ supplied by the DNS
code, and the saddle point of a $Z=114$ nucleus is nearly spherical,
$\alpha_2^{\rm sad}=0.074$ from Eq.~(\ref{eq:saddle}).

\subsection{Energy bookkeeping of the cascade}

At each step the neutron separation energy is recomputed from the FRDM(2012)
mass excesses \cite{Moller2016} rather than kept fixed, since $B_n$ alternates
by more than $1$~MeV between the even-$N$ and the odd-$N$ nuclei of the chain.
The excitation energy of the residual nucleus is
\begin{equation}
E^{*}_{i+1} = E^{*}_{i} - B_n(A_i,Z) - \langle\varepsilon\rangle ,
\label{eq:cascade}
\end{equation}
where $\langle\varepsilon\rangle$ is the average kinetic energy carried away by
the emitted neutron. The temperature entering it is evaluated for the residual
nucleus at the energy already reduced by $B_n$, since it is the residual system
whose level density determines the spectrum of the emitted neutron. Integrating
the Weisskopf spectrum $\varepsilon e^{-\varepsilon/T}$ over the energy which is
actually available, $U = E^{*}_{i}-B_n$, gives
\begin{equation}
\langle\varepsilon\rangle = T\,
\frac{2-(u^{2}+2u+2)e^{-u}}{1-(1+u)e^{-u}},
\qquad u = U/T,
\label{eq:eps}
\end{equation}
which tends to the familiar $2T$ for $u \gg 1$ and never exceeds $U$.

This mean trajectory is used only to evaluate the neutron--fission competition;
the threshold of a channel is set by $P_{xn}$. When the trajectory no longer has
$E^{*}>B_n$, the events which still emit the remaining neutrons are those whose
earlier neutrons carried away less than the average, and $P_{xn}$ already counts
them. At such an energy $E^{*}<B_f$ for every nucleus of this chain, so fission
is closed and the remaining steps contribute a factor of unity. Terminating the
cascade at that point instead, which is the more common choice, cuts each
channel off while its Jackson factor is still of order $0.5$, and leaves
$\Wsur$ ending abruptly at a finite value some $1-5$~MeV above the true
threshold of the channel.

Equations~(\ref{eq:cascade}) and (\ref{eq:eps}) belong to the factorized form of
Eq.~(\ref{eq:wsur}), in which the cascade is followed along a single mean
trajectory. The coupled treatment of the next subsection retains the emission
spectrum which underlies Eq.~(\ref{eq:eps}), but not its reduction to a mean,
and therefore needs neither Eq.~(\ref{eq:cascade}) nor the terminating step
which goes with it.

\subsection{Coupled treatment of the evaporation--fission cascade}
\label{sec:coupled}

The factorization of Eq.~(\ref{eq:wsur}) rests on two separate assumptions. The
first is that the number of neutrons emitted is
decided by $P_{xn}$ alone, independently of the fission competition. The second
is that the branching ratio at the $i$-th step may be evaluated at a single
energy, the one reached along the mean trajectory of Eq.~(\ref{eq:cascade}).

Neither assumption is exact, and they fail in the same direction. Fission
removes the flux preferentially where the excitation energy is high, so that the
nuclei which survive to emit a further neutron are not a fair sample of those
which entered the step: they are colder than the average, and the mean
trajectory therefore overestimates the energy at which the later steps take
place. The emitted neutrons also carry away a distribution of kinetic energies
rather than its mean, and after two or three steps that distribution is broad
enough that the population reaching a given step spans several MeV. Since
$\Gamma_n/\Gamma_f$ varies exponentially with the excitation energy, replacing
the distribution by its mean is not a neutral operation. The error made is
therefore largest for those channels which, at a given excitation energy, lie on
their high-energy falling flank, that is for $x$ below the mean multiplicity of
the cascade. Such residues are fed by the cold tail of the population, the very
part which the mean trajectory cannot represent and which fission has enriched
by removing the hotter members. The channels above the mean multiplicity are
affected in the opposite sense.

These approximations can be removed without leaving the statistical model, by
propagating the whole population instead of a single trajectory. Let $n_x(E)$
denote the population of nuclei which have emitted $x$ neutrons and are left at
the excitation energy $E$, with $n_0(E) = \delta(E-\Ecn)$ at the start. A
nucleus of the chain at the energy $E$ has three exclusive outcomes,
\begin{align}
p_{\rm res}(E) &= \Theta(B_n-E)\,\Theta(B_f-E), \nonumber\\
p_{n}(E)       &= \left[1-\Theta(B_n-E)\right]
                  \frac{\Gamma_n}{\Gamma_n+\Gamma_f}\bigg|_{E},
\label{eq:outcomes}\\
p_{f}(E)       &= 1 - p_{\rm res}(E) - p_{n}(E), \nonumber
\end{align}
namely to remain as an evaporation residue, to emit a further neutron, or to
fission. Here $\Theta$ is the unit step function, equal to unity when its
argument is positive and to zero otherwise, so that the first line selects the
nuclei which are left below the neutron separation energy and can no longer
emit. Such a nucleus is counted as a residue only if it lies below its own
fission barrier as well. The second step function is not redundant: for the
partial waves with $\ell \gtrsim 20\hbar$ the damping $q(\ell)$ of
Eq.~(\ref{eq:damping}) has already pushed $B_f$ below $B_n$, as
Fig.~\ref{fig1} shows, so that a nucleus which stalls at such an angular
momentum is above its barrier and fissions instead of surviving.

The kinetic energy of the emitted neutron is drawn from the Weisskopf spectrum
normalized over the energy which is actually available, $U = E-B_n$,
\begin{equation}
w(\varepsilon;E) = \frac{\varepsilon\,e^{-\varepsilon/T}}
{\int_0^{U}\varepsilon'e^{-\varepsilon'/T}\,d\varepsilon'},
\qquad 0 \leq \varepsilon \leq U,
\label{eq:weisskopf}
\end{equation}
the temperature being that of the residual nucleus, evaluated at $U$. The
population after one further emission is then
\begin{equation}
n_{x+1}(E') = \int_0^{\infty}\!\! dE\; n_x(E)\,p_n(E)\,
w\!\left(E-B_n-E';\,E\right),
\label{eq:transfer}
\end{equation}
and the survival probability of the $xn$ channel is the weight which stops at
exactly $x$ neutrons,
\begin{equation}
\Wsur(\Ecn,x,\ell) = \int_0^{\infty} n_x(E)\,p_{\rm res}(E)\,dE ,
\label{eq:wcoupled}
\end{equation}
the factor $p_{\rm res}$ restricting the integral to $E<B_n$ and, through its
second step function, discarding those nuclei which stall above their own
fission barrier. At $\ell=0$ the second condition is inactive, since $B_f$
exceeds $B_n$ by $0.8-3.5$~MeV for every member of the chain. It becomes active
for $\ell \gtrsim 20\hbar$, where the damping has brought $B_f$ below $B_n$, and
there it is not a small effect: at $\ell=25\hbar$ it lowers $\Wsur$ of the $2n$
channel by a factor which grows from unity at threshold to about $180$ above
$E^{*}=25$~MeV. It is nevertheless retained for correctness rather than for its
numerical weight, since those partial waves contribute so little to
Eq.~(\ref{eq:sigma}) that omitting the condition altogether would leave the
calculated $\sigma_{3n}$ and $\sigma_{4n}$ unchanged to four significant
figures.
No realization probability appears in Eq.~(\ref{eq:wcoupled}). The multiplicity
distribution is generated by the propagation itself, so that $P_{xn}$ of
Eq.~(\ref{eq:pxn}) is no longer needed, and the competition with fission is
evaluated at the energy each nucleus actually has rather than at a mean.

In practice Eqs.~(\ref{eq:outcomes})--(\ref{eq:wcoupled}) are solved on a grid
in excitation energy, of step $0.05$~MeV up to $80$~MeV. Since
Eq.~(\ref{eq:transfer}) is linear in the population, the propagation is carried
out on the identity matrix rather than on one initial energy at a time, so that
a single pass returns $\Wsur$ for every starting energy of the grid at once and
the transfer kernels are built only once. Halving the step changes $\Wsur$ by
a few tenths of a per cent, which is negligible against every other uncertainty
of the model.

Three checks have been made on the solution. With the fission channel switched
off, every compound nucleus must end as some $xn$ residue, and the computed
$\sum_x \Wsur$ equals unity to better than $10^{-9}$, provided the sum is
carried far enough that no open channel is truncated. In the same limit the
multiplicity distribution must reduce to that of a pure evaporation cascade,
and for every channel of appreciable weight it agrees with an independent
Monte-Carlo sampling of the same emission model to better than $5\%$. Finally
each channel opens at the cumulative neutron separation energy $\sum_i B_n$,
and the channels peak in the expected order.

The effect of the coupling is not a small correction. At $\ell=0$ and
$E^{*}=39$~MeV the coupled treatment gives a $3n$ survival probability $17$
times larger than the factorized one, and at $E^{*}=47$~MeV $24$ times larger,
while the $4n$ channel is raised by factors of only $1.7$ and $7.5$ at the same
two energies. The correction is neither uniform nor monotonic in $x$: at
$E^{*}=39$~MeV, where the cascade would emit $4.2$ neutrons on average in the
absence of fission, the factors are $2.8$, $17.3$, $1.7$ and $0.41$ for $x=2$ to
$5$, so that the highest channel is in fact reduced. This is the pattern expected from the
argument given above, the largest correction falling on the channel which is
furthest down its high-energy flank while still carrying appreciable weight. The
coupling therefore acts on the ratio of neighbouring channels rather than on the
overall normalization.

\section{Results and discussion}\label{sec:results}

The method has been applied to $^{292}$Fl produced in the
$^{48}$Ca\,$+\,^{244}$Pu reaction, with the nuclear input collected in
Table~\ref{table1}. The alternation of $B_n$ between $7.05$ and $5.64$~MeV due
to the neutron pairing makes the use of a fixed $B_n$ inadequate, and the shell
correction $\delta W = -9.45$~MeV of $^{292}$Fl is as large as the cold barrier
$B_f^{(0)} \approx 10$~MeV, which confirms that the barrier of these nuclei is of
shell origin.

\begin{table}[htb]
\caption{Nuclear input for the nuclei of the evaporation cascade of $^{292}$Fl.
The neutron separation energy, the shell correction and the ground-state
deformation are taken from FRDM(2012) \cite{Moller2016}. The last two columns
compare the tabulated cold fission barrier of M\"oller \textit{et al.}
\cite{Moller2015} with the value obtained from the LSD macroscopic energy
combined with the topographical theorem, Eq.~(\ref{eq:topo}).}
\begin{ruledtabular}
\begin{tabular}{lccccc}
 & & & & \multicolumn{2}{c}{$B_f^{(0)}$ (MeV)} \\
\cline{5-6}
Nucleus & $B_n$ (MeV) & $\delta W$ (MeV) & $\beta_2$ & table & LSD \\
\hline
$^{292}$Fl & 7.05 & $-9.45$ & $ 0.000$ & 9.98 & 9.45 \\
$^{291}$Fl & 5.64 & $-9.34$ & $ 0.011$ & 9.97 & 9.34 \\
$^{290}$Fl & 7.24 & $-8.95$ & $-0.011$ & 9.89 & 8.95 \\
$^{289}$Fl & 5.96 & $-8.85$ & $-0.021$ & 9.61 & 8.85 \\
$^{288}$Fl & 7.51 & $-8.39$ & $-0.021$ & 9.18 & 8.39 \\
$^{287}$Fl & 6.17 & $-8.41$ & $ 0.064$ & 9.23 & 8.41 \\
$^{286}$Fl & 7.58 & $-7.98$ & $ 0.064$ & 9.00 & 7.98 \\
\end{tabular}\label{table1}
\end{ruledtabular}
\end{table}

\subsection{Confrontation of the entrance channel with the quasifission data}
\label{sec:entrance}

The ER cross section is a product of three factors, and an error made in the
entrance channel can always be compensated by one made in $\Wsur$. Before the
survival probability is examined it is therefore necessary to confront the
first two factors with the data which bear on them directly. For the
$^{48}$Ca\,$+\,^{244}$Pu reaction these are the mass and total kinetic energy
(TKE) distributions of the fissionlike fragments measured by Kozulin
\textit{et al.} \cite{Kozulin2014} with a double-arm time-of-flight
spectrometer, from which a capture cross section and a fusion probability have
been extracted.

It is necessary to state precisely which quantity such a measurement delivers.
The fissionlike events are those which fall inside a gate drawn in the
mass--TKE plane, the products with masses close to those of the projectile and
of the target being excluded as quasielastic and deep-inelastic. The gate of
Fig.~1(c) of Ref.~\cite{Kozulin2014}, which is the one belonging to this
reaction, has been digitised for the present comparison. It extends over the
fragment masses from $A=68.0$ to $224.0$~u and over the total kinetic energies
from $145$ to $295$~MeV, so that it is centred on $A_{\rm CN}/2 = 146.0$~u with
a half width of $78.0$~u; the gates drawn for the other
three reactions of that figure follow the same rule, with half widths of $79$
to $83$~u. The capture cross section quoted with these distributions is not a
directly measured quantity either. The differential cross sections were
measured at one angle, $\theta_{\rm c.m.} \approx 90^{\circ}$, and converted
into $\sigma_{\rm cap}$ under the assumption that the angular distribution is
proportional to $1/\sin\theta_{\rm c.m.}$, the authors stating that neither
detailed angular distributions nor any theory of the angular distribution of
the quasifission fragments are available. What is quoted as $\sigma_{\rm cap}$
is therefore the yield of the gated fissionlike events extrapolated over the
full solid angle under an assumed shape.

The same has to be said, with more force, of the fusion probability. The
quantity extracted in Ref.~\cite{Kozulin2014} is the ratio of the number of
events attributed to the fission of the CN to all the fissionlike events inside
the gate, and it is parameterized there as
\begin{equation}
P_{\rm CN}^{\rm exp}(\Ecm) = \frac{P_0}
{1+\exp\left[c_1\left(c_2-\Ecm/E_{\rm Bass}\right)\right]},
\label{eq:pcnexp}
\end{equation}
with $P_0 = 0.07$, $c_1 = 85$ and $c_2 = 1.002$ for this reaction and
$E_{\rm Bass} = 197.29$~MeV, which corresponds to $E^{*} \simeq 36$~MeV on the
scale of Eq.~(\ref{eq:ecn}). The $P_{\rm CN}$ of Eq.~(\ref{eq:sfus}) is instead
the ratio $\sigma_{\rm fus}/\sigma_{\rm cap}$ taken over the full phase space,
and the two are not the same quantity. The fission of the CN is symmetric and
lies entirely inside the gate, whereas a part of the quasifission yield falls
outside it, so that the denominator of the measured ratio is smaller than the
number of the captured systems and the measured ratio is an upper bound on the
fusion probability. If $f$ denotes the fraction of the quasifission yield which
lies inside the gate, the two are related by
\begin{equation}
P_{\rm CN} = \frac{f\,P_{\rm CN}^{\rm exp}}
{f\,P_{\rm CN}^{\rm exp}+1-P_{\rm CN}^{\rm exp}} .
\label{eq:pcngate}
\end{equation}
The fraction $f$ has been obtained from the charge and mass distributions of
the quasifission fragments given by the same DNS calculation, integrated over
the mass gate quoted above and averaged over the orientations with the same
$\sin\alpha_T\,d\alpha_T$ measure as the partial cross sections themselves,
which gives $f = 0.192$. It is built from the fragment yield of the master
equation at the end of the evolution and therefore carries no explicit
dependence on the bombarding energy; the single value is used at every energy.
The fraction varies strongly with the orientation, from $0.029$ for the two
symmetry axes aligned with the beam to $0.459$ for the near-side configuration,
because the tip-on collisions produce the most asymmetric quasifission.
Corrected in this way, the measured fusion probability falls from
$7.0\times10^{-2}$, which is the saturation value $P_0$ of
Eq.~(\ref{eq:pcnexp}), to $1.4\times10^{-2}$ at the highest energies.

\begin{figure}[htb]
\includegraphics[width=\columnwidth]{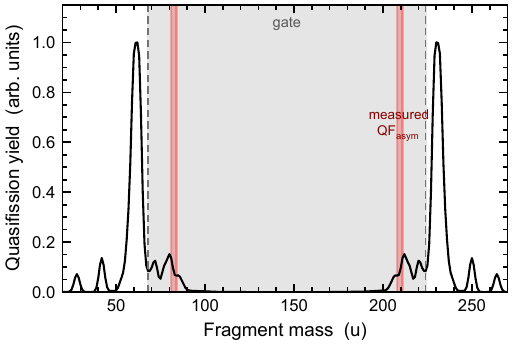}
\caption{Quasifission fragment mass distribution of the
$^{48}$Ca\,$+\,^{244}$Pu reaction calculated with the DNS model, averaged over
the orientations with the $\sin\alpha_T\,d\alpha_T$ measure and folded with the
mass resolution of the spectrometer. The shaded band is the mass gate of
Fig.~1(c) of Ref.~\cite{Kozulin2014}, from $A=68.0$ to $224.0$~u, inside which
the fraction $f = 0.192$ of the yield falls. The narrow bands mark the position
of the asymmetric quasifission peak reported in that work.}
\label{fig:massdist}
\end{figure}

\begin{table}[htb]
\caption{Fusion probability of the $^{48}$Ca\,$+\,^{244}$Pu reaction. The
measured values are those of Fig.~10 of Ref.~\cite{Kozulin2014}, first as
published, which is the ratio formed inside the gate, and then corrected to the
full phase space by Eq.~(\ref{eq:pcngate}) with $f=0.192$; the calculated ones
are the values which enter Eq.~(\ref{eq:sfus}). The comparison is made in
$\Ecm$, which is the variable of that figure and which requires no mass model.}
\begin{ruledtabular}
\begin{tabular}{ccccc}
$\Ecm$ & measured & corrected & this work & ratio \\
(MeV)  & ($10^{-3}$) & ($10^{-3}$) & ($10^{-3}$) & \\
\hline
188.9 &  2.53 &  0.49 &  0.22 & 0.45 \\
193.9 & 17.98 &  3.50 &  3.51 & 1.00 \\
200.5 & 65.51 & 13.28 & 11.67 & 0.88 \\
203.9 & 81.20 & 16.69 & 13.91 & 0.83 \\
\end{tabular}\label{table:pcn}
\end{ruledtabular}
\end{table}

\begin{figure}[htb]
\includegraphics[width=\columnwidth]{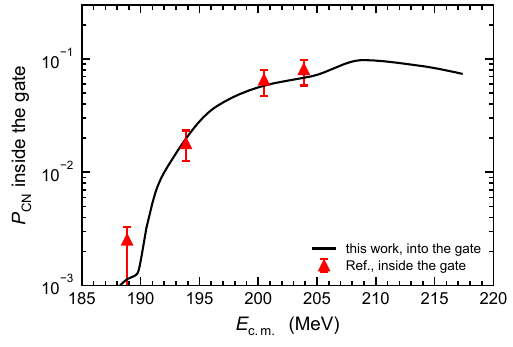}
\caption{Fusion probability of the $^{48}$Ca\,$+\,^{244}$Pu reaction, compared
inside the mass gate of Fig.~\ref{fig:massdist}, which is the quantity the
measurement defines. The triangles are the values of Fig.~10 of
Ref.~\cite{Kozulin2014} as published. The curve is the calculated
$P_{\rm CN}$, a ratio over the full phase space, carried into the gate by the
inverse of Eq.~(\ref{eq:pcngate}),
$P_{\rm CN}^{\rm gate}=P_{\rm CN}/[P_{\rm CN}+f(1-P_{\rm CN})]$, with the same
$f = 0.192$. The comparison is the one of Table~\ref{table:pcn} made in the
other direction, so the ratios are unchanged. The abscissa is $\Ecm$, the
variable of the measurement, which requires no mass model.}
\label{fig:pcn}
\end{figure}

Table~\ref{table:pcn} compares the corrected measurement with the calculated
$P_{\rm CN}$, and Fig.~\ref{fig:pcn} shows the same comparison carried out
inside the gate instead, where the measurement is defined. The four measured points correspond to the
beam energies $E_{\rm lab}=226$, $232$, $240$ and $244$~MeV quoted in Table~I
of Ref.~\cite{Kozulin2014}, and the two highest of them fall within $1$~MeV of
points of the DNS grid, so that no interpolation is needed. At those two
energies the calculated fusion probability is $0.88$ and $0.83$ of the
corrected measurement, and at $\Ecm = 193.9$~MeV the two agree. The order of
magnitude which separates the calculated fusion probability from the published
one is therefore not a disagreement but a difference of definition between a
ratio formed inside a gate and a ratio formed over the whole phase space. Only
the lowest measured point, which lies close to the threshold where the
calculated $\sigma_{\rm fus}$ falls by three orders of magnitude over a few MeV,
is not reproduced; the calculation gives $0.45$ of the corrected value there,
and the steepness of the excitation function in that region makes the
comparison sensitive to a fraction of an MeV in the energy scale. It should be
noted that $P_{\rm CN}$ enters Eq.~(\ref{eq:sigma}) through
$\sigma_{\rm fus}$ as a factor common to the two channels, so that a
disagreement of this size moves both excitation functions together and cannot
by itself produce the ratio $\sigma_{4n}/\sigma_{3n}$ discussed below. The
capture cross section itself is not tested here, since the published value
rests on the angular-distribution assumption quoted above.

Three reservations have to be attached to this correction, and
Fig.~\ref{fig:massdist} makes them visible. The first is that $f$ is taken from
the calculated mass distribution and is sensitive to the position of the edge of
the gate: a lower edge at $65$, $68$ and $70$~u gives $f = 0.219$, $0.192$ and
$0.170$. The reason is apparent in the figure, where the calculated quasifission
yield peaks at a light-fragment mass of $63$~u, only $5$~u below the edge, so
that the gate cuts the distribution where it is steepest. The second is that
only the mass part of the gate has been applied and not its TKE part, so that
$f$ is an upper limit and the corrected $P_{\rm CN}$ of Table~\ref{table:pcn} is
an upper limit as well. The third concerns the calculated distribution itself:
its heavy component peaks at $A \simeq 229$~u, whereas the asymmetric
quasifission peak of Ref.~\cite{Kozulin2014} lies at $208$ to $211$~u, where it
is driven by the $Z=82$ and $N=126$ shells. The calculated distribution is thus
about $20$~u too asymmetric in the heavy fragment, and the correction is only as
good as the distribution from which $f$ is obtained. This last point is a failure of the calculated mass distribution
and not of the correction procedure, and it is the reason why
Table~\ref{table:pcn} should be read as a consistency check of the entrance
channel and not as a validation of it.

\subsection{Fission barrier and survival probability}

\begin{figure}[htb]
\includegraphics[width=\linewidth]{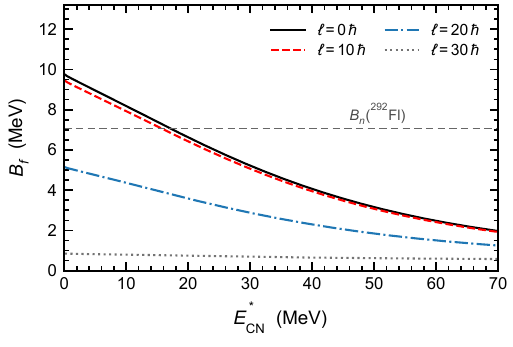}
\caption{Fission barrier of $^{292}$Fl as a function of the excitation
energy for four values of the angular momentum, from Eqs.~(\ref{eq:bf}) and
(\ref{eq:damping}); the dashed line is the neutron separation energy of
$^{292}$Fl.}\label{fig1}
\end{figure}

Figure~\ref{fig1} shows the barrier of $^{292}$Fl given by Eqs.~(\ref{eq:bf})
and (\ref{eq:damping}). At $\ell=0$ it falls from $9.79$~MeV at
$E^{*}\simeq 0$ to $2.48$~MeV at $E^{*}=60$~MeV, crossing the neutron
separation energy near $E^{*}\approx 17$~MeV. The dependence on the angular
momentum is much sharper: at $E^{*}=35$~MeV the barrier is $4.61$~MeV at
$\ell=0$ but only $0.67$~MeV at $\ell=30\hbar$, because $q(\ell)$ removes the
shell contribution over a few $\hbar$ around $\ell_{1/2}=20\hbar$. The residues
are therefore formed almost entirely by the low partial waves, and the sum in
Eq.~(\ref{eq:sigma}) is in practice restricted to $\ell \lesssim 20\hbar$.

The survival probability which follows from this barrier is shown in
Fig.~\ref{fig2}(b) for the two channels which have been measured. Each opens at
its own threshold, rises within a few MeV to a maximum and then falls close to
exponentially, the $3n$ channel losing an order of magnitude every $3.8$~MeV
and the $4n$ channel every $5.4$~MeV. The fall is a direct consequence of the
damping of Fig.~\ref{fig1}: every additional MeV of excitation energy both
lowers the barrier and increases the fission width through the exponent of
Eq.~(\ref{eq:gnf}).

\begin{figure}[htb]
\includegraphics[width=\linewidth]{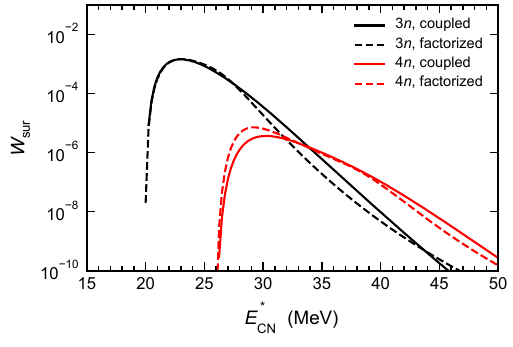}
\caption{Survival probability of the $3n$ and $4n$ channels at $\ell=0$,
obtained from the coupled cascade of Sec.~\ref{sec:coupled} (solid) and from
the factorized form of Eq.~(\ref{eq:wsur}) (dashed). Both are computed with the
tabulated barrier of M\"oller \textit{et al.} \cite{Moller2015}, so that the
two line styles of a given colour differ only in the treatment of the
cascade.}\label{figcf}
\end{figure}

\begin{figure*}[htb]
\includegraphics[width=0.95\linewidth]{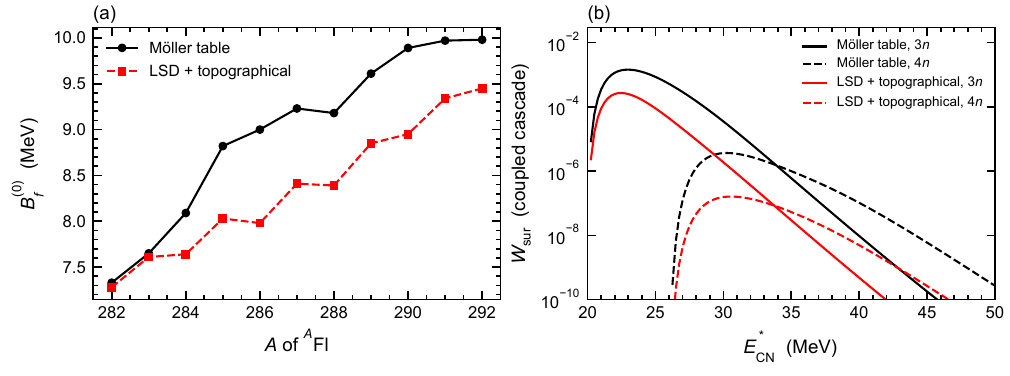}
\caption{(a) Cold fission barrier of the Fl isotopes from the table of
M\"oller \textit{et al.} \cite{Moller2015} and from the LSD macroscopic energy
combined with the topographical theorem, Eq.~(\ref{eq:topo}). (b) Survival
probability of the $3n$ and $4n$ channels at $\ell=0$ obtained from the coupled
cascade with each of the two prescriptions: colour distinguishes the barrier,
line style the channel. Every other ingredient is identical between the two
colours, so their vertical separation measures the sensitivity of $\Wsur$ to
$B_f$ alone.}\label{fig2}
\end{figure*}

Figure~\ref{figcf} compares the survival probability obtained from the coupled
cascade with that of the factorized form, both computed with the tabulated
barrier so that the two differ only in the treatment of the cascade. They agree
near the maximum of each channel and part company above it. For the $3n$
channel the coupled result exceeds the factorized one by a factor of $15$ to
$30$ over $E^{*}=35-50$~MeV. For the $4n$ channel the effect is much smaller
near the maximum, a factor of $1.3$ at $E^{*}=35$~MeV, but it grows with energy
and reaches $15$ at $E^{*}=50$~MeV;
below the maximum the ordering reverses and the factorized form is the larger
of the two. The divergence is therefore concentrated on the falling flank,
which is where the neglected correlation between the multiplicity and the
survival against fission matters most, and it is the reason the two treatments
give excitation functions of visibly different width in Figs.~\ref{fig3} and
\ref{fig4}.

\begin{table*}[htb]
\caption{Separation of the two effects at $\ell=0$. Column (1) is the factorized
form of Eq.~(\ref{eq:wsur}); column (2) replaces the Jackson multiplicity by
$P^{(0)}_{x}$, the multiplicity of the same evaporation kernel without fission,
keeping the branching ratios along the mean trajectory; column (3) is the
coupled result. The step (1)$\rightarrow$(2) is the change of multiplicity
prescription alone and the step (2)$\rightarrow$(3) the replacement of the mean
trajectory by the average over the histories which end in the channel. The
multiplicity prescription accounts for none of the enhancement.}
\begin{ruledtabular}
\begin{tabular}{ccccccc}
Channel & $E^{*}$ (MeV) & (1) & (2) & (3) & (2)/(1) & (3)/(2) \\
\hline
$3n$ & 39 & $3.12\times10^{-9}$  & $1.04\times10^{-9}$  & $7.72\times10^{-8}$  & 0.33 & 74  \\
$3n$ & 47 & $2.91\times10^{-11}$ & $4.09\times10^{-12}$ & $8.25\times10^{-10}$ & 0.14 & 202 \\
$4n$ & 39 & $4.62\times10^{-8}$  & $4.26\times10^{-8}$  & $8.68\times10^{-8}$  & 0.92 & 2.0 \\
$4n$ & 50 & $3.60\times10^{-11}$ & $1.17\times10^{-11}$ & $5.36\times10^{-10}$ & 0.33 & 46  \\
\end{tabular}\label{table:control}
\end{ruledtabular}
\end{table*}

It has to be established that this difference is caused by the correlation
itself and not by the two different descriptions of the neutron multiplicity
which the two treatments happen to use, the Jackson expression of
Eq.~(\ref{eq:pxn}) in one case and the evaporation kernel of
Eq.~(\ref{eq:eps}) in the other. The multiplicity distribution generated by that
kernel with the fission width switched off, $P^{(0)}_{x}$, isolates the two
effects. Table~\ref{table:control} compares three quantities: the factorized
form as published, the same product with $P_{xn}$ replaced by $P^{(0)}_{x}$, and
the coupled result. Replacing the multiplicity prescription alone makes the
survival probability \emph{smaller}, by a factor of $3$ at $E^{*}=39$~MeV and of
$7$ at $47$~MeV, so that it acts against the enhancement rather than producing
it; the remaining step raises it by factors of $74$ and $202$. The whole of the difference, and more, comes from the last step, in which
the branching ratios evaluated along one mean trajectory are replaced by their
average over the histories which actually end in the channel. The reason is
visible in the trajectory itself: the $xn$ channel requires the residue to stop
below the neutron separation energy of the residual nucleus, $5.96$~MeV for
$^{289}$Fl, whereas the mean trajectory of the $3n$ product still carries
$12.7$~MeV at $E^{*}=39$~MeV. The branching ratios of the factorized form are
therefore evaluated on histories which are not $3n$ events, and which are hotter,
and hence more fissile, than those which are. For the $4n$ channel the mean
trajectory does stop inside its own channel up to $E^{*}\simeq40$~MeV, which is
why the factorization is adequate there and fails only on the upper flank.

The evaporation kernel of Eq.~(\ref{eq:eps}) is itself an approximation, the
temperature being evaluated once at $U=E^{*}-B_n$ rather than following the
level density of the daughter at $U-\varepsilon$. Since the effect described
above works through the low-energy tail of the surviving population, the
calculation has been repeated with the exact kernel
$w(\varepsilon)\propto\varepsilon\,\rho_{d}(U-\varepsilon)$, the inverse cross
section being left out because the Dostrovsky parameter for neutrons on
$A\simeq291$ is $\beta=-0.0016$~MeV, so that $\sigma_{\rm inv}$ is flat to
$0.3\%$ over the relevant range. With $T_0$ refitted for each kernel the exact
kernel gives the poorer description of the measured excitation functions, a
$\chi^{2}$ per point of $4.2$ against $2.1$ for Eq.~(\ref{eq:eps}). It does move
the ratio of the neighbouring channels, giving $5.6$ at $E^{*}=39$~MeV against
$0.9$; as noted above the measured ratio at that energy lies between $1.1$ and
$4.2$ depending on which of the two inconsistent $3n$ measurements is used, so
that this observable alone does not discriminate between the two kernels, but
the fit over all fourteen points does. The exact kernel puts more energy into
each emitted neutron and so cools the residue faster, which shortens the cascade
and steepens both excitation functions beyond what the data support. The results
quoted throughout are those of Eq.~(\ref{eq:eps}), which is retained as an
explicit approximation rather than for its convenience.

The maxima of the two channels differ by much more than the cross sections
which follow from them, and the reason is visible in Fig.~\ref{fig2}(b). The
maximum of $\Wsur$ is $7.2\times10^{-4}$ for the $3n$ channel and
$1.1\times10^{-6}$ for the $4n$ one, a factor of about $700$, but the two are
reached $7.6$~MeV apart, at $E^{*}=23.1$ and $30.6$~MeV, so that a comparison of
the two heights is not a comparison at the same excitation energy. At a fixed
$E^{*}$ the channels lie much closer together, and the two curves cross at
$E^{*}=38.4$~MeV, above which the $4n$ channel is the larger of the two. The
measured residues all lie above $E^{*}=31$~MeV, which is far down the falling
flank of the $3n$ channel but close to the maximum of the $4n$ one.

The partial fusion cross section which multiplies $\Wsur$ in
Eq.~(\ref{eq:sigma}) is the same for both channels. It does not cancel from
their ratio identically, since the sum over partial waves weights $\Wsur$
differently for the two channels, but it very nearly does: tilting the weights
$\Delta\ell\,\sigma_{\rm fus}(\Ecm,\ell)$ by $\pm30\%$ across the surviving
range, or truncating them at $\ell=25\hbar$ or $15\hbar$, changes
$\sigma_{4n}/\sigma_{3n}$ by less than $1\%$, because the surviving waves span a
narrow interval over which the two survival probabilities have a similar
dependence on $\ell$. To that accuracy $\sigma_{4n}/\sigma_{3n}$ follows the
ratio of the survival probabilities at equal energy rather than the ratio of
their maxima. The two
orders of magnitude which separate the maxima in Fig.~\ref{fig2}(b) are
therefore already spent before the measured interval is reached, and the two
excitation functions are of comparable height.

The displacement of the maxima is itself a consequence of the propagation. Each
further neutron costs about two orders of magnitude in $\Wsur$ at a fixed
energy, and at the same time moves the maximum of the channel up in energy by
the neutron separation energy together with the mean kinetic energy carried
away. Above its own maximum a channel is suppressed twice over: fission removes
a part of the population, and the part which survives no longer stops after $x$
neutrons but goes on to emit an $(x+1)$-th. Both mechanisms are present in the
factorized form as well, the second through the term of Eq.~(\ref{eq:pxn})
built on $B_{x+1}$; what the factorized form cannot represent is that the two
are correlated, since the histories which emit the $(x+1)$-th neutron are also
those which stayed hot enough to fission.

\subsection{Sensitivity to the barrier prescription}

The last two columns of Table~\ref{table1} compare the two prescriptions for the
cold barrier, and Fig.~\ref{fig2}(a) extends the comparison over the Fl
isotopes for which both are available. The two agree to better than $0.1$~MeV
for $A \leq 283$, but the tabulated barrier is systematically higher for the
heavier isotopes, by $0.53$~MeV for $^{292}$Fl and by as much as $0.94$~MeV for
$^{290}$Fl. This difference can be understood. The topographical theorem replaces
the microscopic correction at the saddle point by the one evaluated at the
ground state, which is a good approximation only when the saddle is far enough
from the region where the ground-state shell structure is built up. For the
nuclei considered here, whose saddle is nearly spherical, that condition is
not well satisfied, and the
tabulated barrier of Ref.~\cite{Moller2015}, which follows the potential energy
over the whole deformation space, is expected to be the more reliable of the
two.

Because the same damping is applied in both cases, this offset is independent of
$E^{*}$ and of $\ell$. It cannot, however, be identified with the microscopic
correction at the saddle point alone. The two prescriptions do not share a
common macroscopic reference: one is the tabulated microscopic--macroscopic
barrier of Ref.~\cite{Moller2015}, obtained from an explicit multidimensional
saddle search, the other combines the LSD macroscopic energy with the
topographical theorem, which sets the saddle correction to zero. Their
difference therefore contains both the residual saddle correction and whatever
the two macroscopic energies differ by at the saddle, and the present comparison
cannot separate the two. What can be said is that the offsets of
Table~\ref{table1}, $0.5-0.9$~MeV in this chain and always in the sense of a
higher tabulated barrier, are of the size and of the direction anticipated in
Ref.~\cite{Lublin}, where the barriers obtained with the theorem were found to
lie systematically below the measured ones for $Z>70$.

It is important to note how strongly such a sub-MeV difference propagates.
Figure~\ref{fig2}(b) shows the survival probability of the $3n$ and $4n$
channels obtained from the coupled cascade with each prescription. The
lower LSD barrier increases the
fission width through the exponent of Eq.~(\ref{eq:gnf}) at every step of the
cascade, and $\Wsur$ is reduced by a factor of $14$ to $24$ over the whole
interval $E^{*}=30-50$~MeV. The two effects are independent: the coupling
changes the ratio of neighbouring channels, the barrier changes the absolute
scale of both. Since the fusion cross section of Eq.~(\ref{eq:sigma}) is common
to the two prescriptions, the same factor is carried into $\sigma_{xn}$. The barrier of the nuclei
of the cascade therefore has to be known to better than a few hundred keV before
the absolute value of $\sigma_{xn}$ can be predicted, which is a stronger
requirement than that placed on any other ingredient of the model.

These considerations should not be read as establishing that one of the two
prescriptions is correct. Neither barrier in this region has been measured, and
the tabulated barriers are themselves model dependent: they are static, they
correspond to zero angular momentum, and they depend on the deformation space in
which the least-energy path is sought. For the superheavy nuclei the results of
different microscopic--macroscopic calculations
\cite{Moller2015,Kowal2010,Jachimowicz2021} differ by $1-2$~MeV, that is by more
than the offset between the two prescriptions compared here. What the comparison
does establish is the sensitivity: an uncertainty of $1$~MeV in $B_f$, which is
the present state of the art for these nuclei, translates into more than an
order of magnitude in the predicted ER cross section, and this, rather than the
statistical model itself, is what limits the accuracy of such predictions.

\subsection{Evaporation-residue cross sections}
\label{sec:er}

\begin{figure}[htb]
\includegraphics[width=\linewidth]{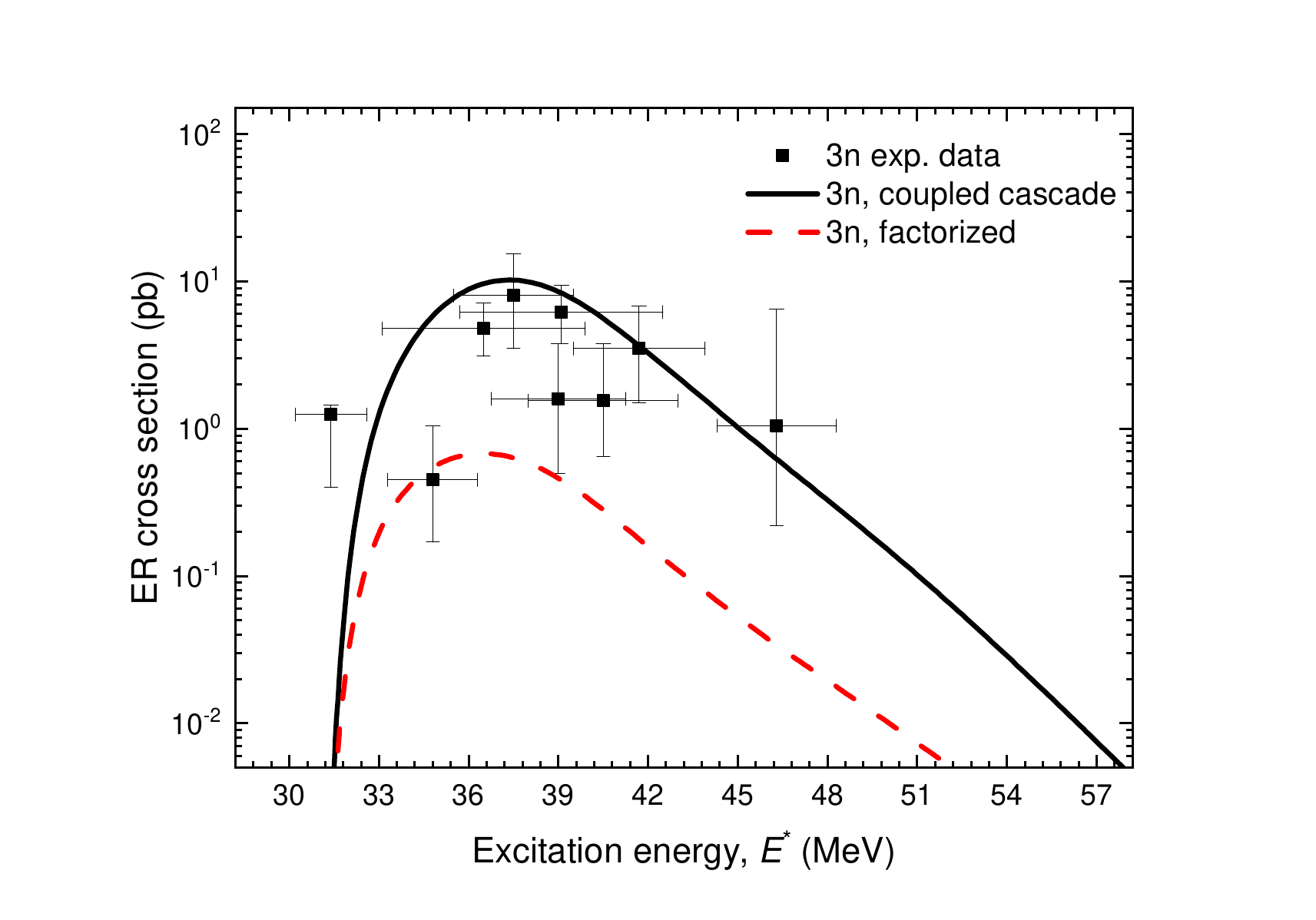}
\caption{Calculated and measured evaporation-residue cross section of the $3n$
channel of the $^{48}$Ca\,$+\,^{244}$Pu reaction, leading to $^{289}$Fl. The
solid curve is obtained from the coupled cascade of Sec.~\ref{sec:coupled}, the
dashed one from the factorized form of Eq.~(\ref{eq:wsur}), all other ingredients
being identical. The curves are drawn through the calculated points, which lie
on the energy grid of the DNS output. The data are those of
Refs.~\cite{Oganessian2004b,Oganessian2004a,Dullmann2010,Gates2011,SamarkRoth2023,Gates2026}.}
\label{fig3}
\end{figure}

\begin{figure}[htb]
\includegraphics[width=\linewidth]{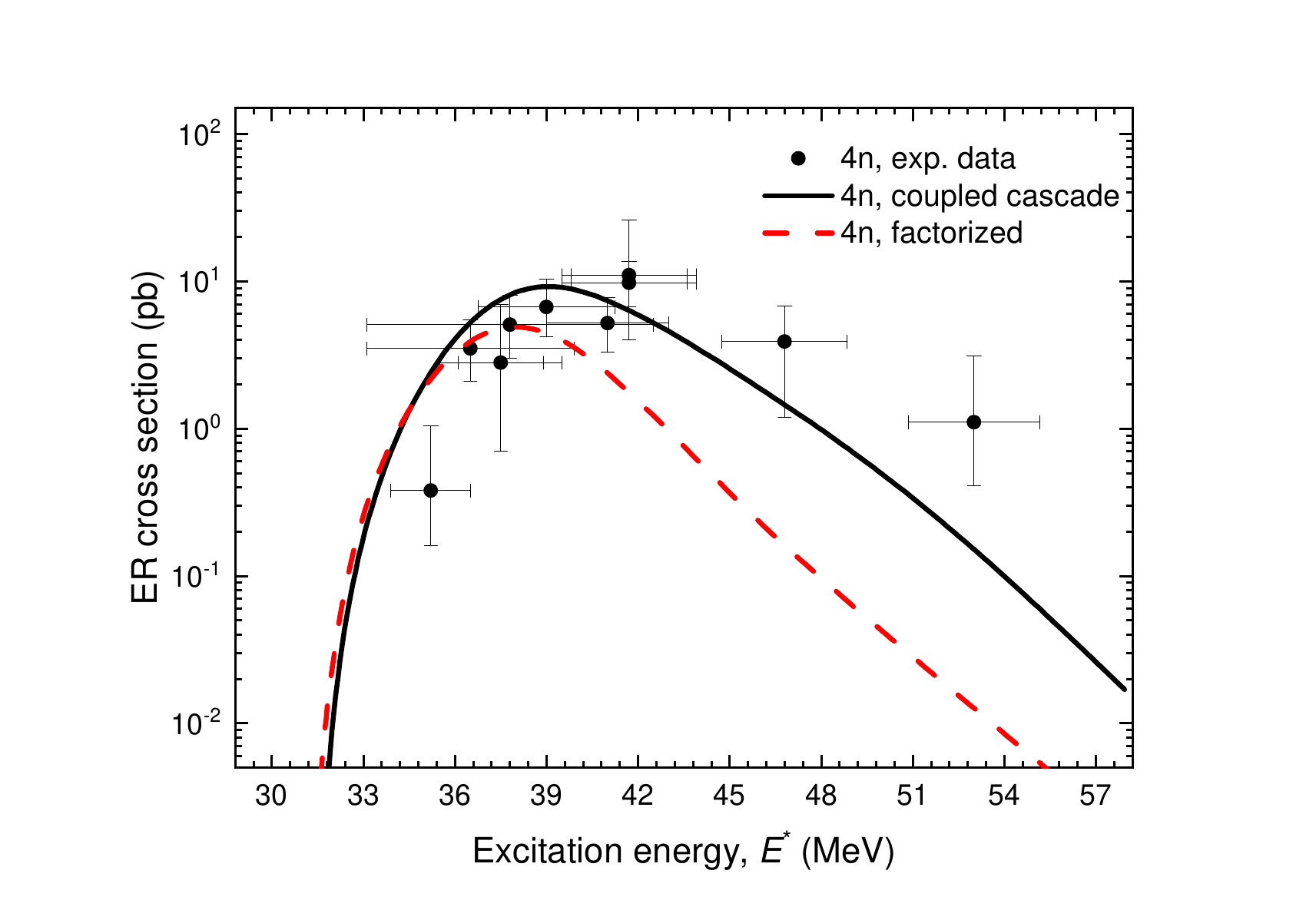}
\caption{The same as in Fig.~\ref{fig3}, but for the $4n$ channel, leading to
$^{288}$Fl.}\label{fig4}
\end{figure}

The excitation functions obtained from Eq.~(\ref{eq:sigma}) are compared in
Figs.~\ref{fig3} and \ref{fig4} with all the $^{289}$Fl $(3n)$ and $^{288}$Fl
$(4n)$ cross sections measured for this reaction at FLNR
\cite{Oganessian2004b,Oganessian2004a}, at GSI
\cite{Dullmann2010,Gates2011,SamarkRoth2023} and at LBNL \cite{Gates2026}. The
comparison at the measured energies is collected in Table~\ref{table2}.

Both excitation functions rise steeply from below, reach a maximum and fall.
The $3n$ channel peaks at $17.5$~pb and the $4n$ channel at $13.3$~pb, both at
$E^{*}=37.1$~MeV, so that the two maxima are of comparable height and nearly
coincident in energy. Both lie within the interval covered by the
measurements.

Over the measured points the geometric mean of the calculated-to-experimental
ratio is $1.75$ and its geometric spread, defined as the exponential of the
sample standard deviation of $\ln(\sigma_{\rm calc}/\sigma_{\rm exp})$, is a
factor of $3.7$; six of the fourteen points are reproduced within a factor of
two. Separated by channel the means are $2.94$ for $3n$ and $1.04$ for $4n$, so
that the $4n$ channel is reproduced on average while the $3n$ channel is
overestimated by about a factor of three. This count of points within a factor of
two is a descriptive statement and not a goodness-of-fit measure.

The description is closest between $E^{*}=39$ and $42$~MeV, where the $4n$
cross section is reproduced within a factor of $1.7$ at every point and the
$3n$ cross section within a factor of $1.8$ at the TASISpec point at
$E^{*}=39.1$~MeV and at the TASCA point at $E^{*}=41.7$~MeV. The largest
departures are on the rising flank, where the FLNR point at $E^{*}=34.8$~MeV and
the LBNL one at $E^{*}=39.0$~MeV are exceeded by factors of $29$ and $7$. It should be noted that the two $3n$
measurements at $39.0$ and $39.1$~MeV differ from each other by a factor of
$3.9$; their quoted intervals do meet, at about $3.8$~pb, so that a calculation
passing through that value is consistent with both, but no calculation can
reproduce both central values.

\begin{table*}[htb]
\caption{Calculated and measured evaporation-residue cross sections of the
$^{48}$Ca\,$+\,^{244}$Pu reaction. The last row of each block is a
non-observation upper limit and is excluded from the mean ratio quoted in the
text. Laboratories: F = FLNR/DGFRS, T = GSI/TASCA, S = GSI/TASISpec,
L = LBNL/SHREC.}
\begin{ruledtabular}
\def\arraystretch{1.35}%
\begin{tabular}{clrr@{\qquad}clrr}
\multicolumn{4}{c}{$3n$ channel} & \multicolumn{4}{c}{$4n$ channel} \\
\cline{1-4}\cline{5-8}
Lab. & $E^{*}$ & $\sigma_{\rm exp}$ & $\sigma_{\rm calc}$ &
Lab. & $E^{*}$ & $\sigma_{\rm exp}$ & $\sigma_{\rm calc}$ \\
     & (MeV)   & (pb) & (pb) & & (MeV) & (pb) & (pb) \\
\hline
F & 34.8 & $0.45^{+0.60}_{-0.28}$ & 12.84 & T & 37.5 & $2.80^{+4.20}_{-2.10}$ & 13.20 \\
T & 37.5 & $8.00^{+7.40}_{-4.50}$ & 17.05 & S & 37.8 & $5.10^{+2.90}_{-2.10}$ & 12.99 \\
L & 39.0 & $1.60^{+2.20}_{-1.10}$ & 11.39 & L & 39.0 & $6.70^{+3.60}_{-2.50}$ & 11.50 \\
S & 39.1 & $6.20^{+3.20}_{-2.40}$ & 10.96 & F & 41.0 & $5.20^{+2.60}_{-1.90}$ &  8.29 \\
F & 40.5 & $1.55^{+2.25}_{-0.90}$ &  6.10 & T & 41.7 & $9.80^{+3.90}_{-3.10}$ &  7.24 \\
T & 41.7 & $3.50^{+3.30}_{-2.00}$ &  3.85 & F & 46.8 & $3.90^{+2.90}_{-2.70}$ &  1.48 \\
F & 46.3 & $1.05^{+5.45}_{-0.83}$ &  0.61 & F & 53.0 & $1.11^{+2.00}_{-0.70}$ &  0.16 \\
F & 31.4 & $<1.25$                &  0.01 & F & 35.2 & $<0.38$               &  7.34 \\
\end{tabular}\label{table2}
\end{ruledtabular}
\end{table*}

Three discrepancies remain. The first is the ratio of the two channels,
$\sigma_{4n}/\sigma_{3n}=1.01$ at $E^{*}=39$~MeV. The measured value there is
not well determined: taking the LBNL $3n$ point gives $4.2$ and the TASISpec one
gives $1.1$, and those two measurements are the mutually inconsistent pair just
discussed, the $3n$ assignment of the former resting on two tentative chains.
The calculated ratio sits at the lower end of that range. The second is the
high-energy tail, which is still too steep: the $3n$ cross section at
$E^{*}=46.3$~MeV and the $4n$ cross section at $E^{*}=46.8$~MeV are
underestimated by factors of $1.7$ and $2.6$, and the $4n$ cross section at
$E^{*}=53$~MeV by a factor of $7$. The third is the $4n$ non-observation limit
at $E^{*}=35.2$~MeV, which is exceeded by a factor of $19$. This last is the
most serious of the three. It lies on the rising flank of that channel, where
the calculated cross section varies most rapidly with energy, and it is the same
region in which the $3n$ channel is overestimated by a factor of $29$ at
$E^{*}=34.8$~MeV; no value of $T_0$ was found which satisfies the limit without
destroying the description at the maximum. The $3n$ limit at $E^{*}=31.4$~MeV is
satisfied by more than two orders of magnitude.

The Kramers and Strutinsky factors of Eqs.~(\ref{eq:kramers}) and
(\ref{eq:strutinsky}) act on the slope of the excitation function rather than on
its normalization, since their product grows with the temperature, and neither
of them introduces an adjustable parameter. The damping temperature $T_0$ is
adjusted to the data. The quantity minimized is the $\chi^{2}$ formed in the
logarithm of the cross section, $\sum[\ln(\sigma_{\rm calc}/\sigma_{\rm
exp})/\sigma_{\ln}]^{2}$ with
$\sigma_{\ln}=\tfrac12\ln[(\sigma+\Delta^{+})/(\sigma-\Delta^{-})]$, taken over
the fourteen measured points of Table~\ref{table2}; the logarithm is used
because the cross sections span three decades and the errors are strongly
asymmetric, and a $\chi^{2}$ formed in the cross section itself is minimized by
a systematic underestimate. The two non-observation limits are not included in
the fit and are reported separately. The value obtained is $T_0=1.20$~MeV,
against the $1.16$~MeV used with the same prescription in
Refs.~\cite{Giardina2018,Mandaglio2018}; the minimum is shallow, values from
$1.20$ to $1.24$~MeV lying within one unit of $\chi^{2}$ per point.

\section{Summary and Conclusion}\label{sec:conclusion}

The Vandenbosch--Huizenga treatment of the neutron--fission competition has been
combined with the dinuclear system model of the entrance channel and applied to
$^{292}$Fl formed in the $^{48}$Ca\,$+\,^{244}$Pu reaction. The level density is
taken from the four-term expression with the Ignatyuk energy dependence, and the
fission barrier is damped in both the temperature and the angular momentum. All
the nucleus-dependent quantities are taken from the FRDM(2012) tables for every
nucleus of the cascade.

The usual factorization of the survival probability into a realization
probability of the $xn$ channel and a product of branching ratios along a mean
trajectory has been replaced by a coupled treatment, in which the population of
the cascade is propagated over the excitation energy and the neutron--fission
competition is evaluated at the energy each nucleus actually has. The
factorization was found to be a poor approximation for these nuclei. It
underestimates the $3n$ survival probability by a factor of $25$ at
$E^{*}=39$~MeV and of $28$ at $E^{*}=47$~MeV, while affecting the $4n$ channel
much less near its maximum, so that its principal effect is on the ratio of
neighbouring channels. A control in which the multiplicity of the coupled
kernel is substituted into the factorized product shows that the change of
multiplicity prescription accounts for none of this difference and acts in the
opposite sense, so that all of it comes from replacing the branching ratios
along a mean trajectory by their average over the histories which end in the
channel. The coupled solution conserves probability and reproduces an
independent Monte-Carlo cascade in the no-fission limit.

The calculations show that above the maximum of a given channel $\Wsur$ of
$^{292}$Fl loses an order of magnitude every $5.4$~MeV for the $3n$ channel and
every $7.9$~MeV for the $4n$ one, and that the angular momentum dependence of
the barrier is much sharper
than the temperature dependence, so that the residues are formed almost
entirely by the partial waves with $\ell \lesssim 20\hbar$. Over the measured
points the geometric mean of the calculated-to-experimental ratio is $1.75$
with a spread of a factor of $3.7$, and six of the fourteen points fall within
a factor of two.

This description has been obtained with the Kramers and Strutinsky corrections
to the fission width, neither of which adds an adjustable parameter, both acting
on the slope of the excitation function rather than on its normalization. The
damping temperature is fitted to the data by minimizing the $\chi^{2}$ formed in
the logarithm of the cross section over the fourteen measured points, and the
value obtained, $1.20$~MeV, is to be compared with the value used with the same
prescription in Refs.~\cite{Giardina2018,Mandaglio2018}.

The comparison leaves three open points. The ratio $\sigma_{4n}/\sigma_{3n}$ is
$1.01$ at $E^{*}=39$~MeV, at the lower end of the range from $1.1$ to $4.2$
which the two mutually inconsistent $3n$ measurements at that energy allow. The
high-energy tail is still underestimated, by a factor of $7$ at
$E^{*}=53$~MeV. And the cross sections are too large on the rising flank: the
$3n$ channel is overestimated by a factor of $29$ at $E^{*}=34.8$~MeV, and the
$4n$ non-observation limit at $E^{*}=35.2$~MeV is exceeded by a factor of $19$.
The last of these is the most serious disagreement of the present calculation,
and no value of the damping temperature was found which removes it while
preserving the description at the maximum.

The comparison of the two barrier prescriptions shows how demanding this
observable is on the nuclear input. The tabulated barriers of M\"oller
\textit{et al.} and the topographical estimate differ by less than $1$~MeV for
the nuclei of the cascade, and yet the survival probability obtained with them
differs by more than an order of magnitude. Any statement about the absolute
value of the ER cross section is therefore limited by the accuracy of $B_f$
alongside the treatment of the cascade itself, which is shown here to be worth a
comparable factor. The presented method may serve as a basis for
such an analysis of the reactions leading to the neighbouring superheavy
nuclei.

\section*{Acknowledgements}

The author is grateful to A. K. Nasirov for valuable discussions.

\bibliographystyle{apsrev4-2}
\bibliography{references}

@article{Hofman2000,
  title = {The discovery of the heaviest elements},
  author = {Hofmann, S. and M\"unzenberg, G.},
  journal = {Rev. Mod. Phys.},
  volume = {72},
  issue = {3},
  pages = {733--767},
  numpages = {45},
  year = {2000},
  month = {Jul},
  publisher = {American Physical Society},
  doi = {10.1103/RevModPhys.72.733},
  url = {https://link.aps.org/doi/10.1103/RevModPhys.72.733}
}

@article{Oganessian2007,
doi = {10.1088/0954-3899/34/4/R01},
url = {https://dx.doi.org/10.1088/0954-3899/34/4/R01},
year = {2007},
month = {mar},
publisher = {},
volume = {34},
number = {4},
pages = {R165},
author = {Yuri Oganessian},
title = {Heaviest nuclei from 48Ca-induced reactions},
journal = {Journal of Physics G: Nuclear and Particle Physics}
}

@article{Dullmann2010,
  title = {Production and Decay of Element 114: High Cross Sections and the New Nucleus
  $^{277}\mathrm{Hs}$},
  author = {D\"ullmann, Ch. E. and others},
  journal = {Phys. Rev. Lett.},
  volume = {104},
  issue = {25},
  pages = {252701},
  numpages = {5},
  year = {2010},
  month = {Jun},
  publisher = {American Physical Society},
  doi = {10.1103/PhysRevLett.104.252701},
  url = {https://link.aps.org/doi/10.1103/PhysRevLett.104.252701}
}

@article{Antonenko1995,
  title = {Compound nucleus formation in reactions between massive nuclei: Fusion barrier},
  author = {Antonenko, N. V. and Cherepanov, E. A. and Nasirov, A. K. and Permjakov, V. P. and Volkov,
  V. V.},
  journal = {Phys. Rev. C},
  volume = {51},
  issue = {5},
  pages = {2635--2645},
  numpages = {0},
  year = {1995},
  month = {May},
  publisher = {American Physical Society},
  doi = {10.1103/PhysRevC.51.2635},
  url = {https://link.aps.org/doi/10.1103/PhysRevC.51.2635}
}

@article{Adamian1997,
  title = {Friction coefficient for deep-inelastic heavy-ion collisions},
  author = {Adamian, G. G. and Jolos, R. V. and Nasirov, A. K. and Muminov, A. I.},
  journal = {Phys. Rev. C},
  volume = {56},
  issue = {1},
  pages = {373--380},
  numpages = {0},
  year = {1997},
  month = {Jul},
  publisher = {American Physical Society},
  doi = {10.1103/PhysRevC.56.373},
  url = {https://link.aps.org/doi/10.1103/PhysRevC.56.373}
}

@article{Nasirov2005,
title = {The role of orientation of nucleus symmetry axis in fusion dynamics},
author = {Avazbek Nasirov and Akira Fukushima and Yuka Toyoshima and Yoshihiro Aritomo and Akhtam
Muminov and Shuhrat Kalandarov and Ravshanbek Utamuratov},
journal = {Nucl. Phys. A},
volume = {759},
number = {3},
pages = {342-369},
year = {2005},
issn = {0375-9474},
doi = {https://doi.org/10.1016/j.nuclphysa.2005.05.152},
url = {https://www.sciencedirect.com/science/article/pii/S0375947405008572}
}

@article{Nasirov2019,
  title = {The effect of the neutron and proton numbers ratio in colliding nuclei on the formation of
  the evaporation residues in the $^{34}$S + $^{208}$Pb
and $^{36}$S + $^{206}$Pb reactions},
  author = {Nasirov, A. K. and  Kayumov, B. M. and  Mandaglio, G. and  Giardina, G.
   and Kim, K. and  Kim, Y.},
  journal = {European Physical Journal A},
  volume = {55},
  issue = {2},
  pages = {29},
  numpages = {15},
  year = {2019},
  month = {Feb},
  publisher = {Springer},
  doi = {10.1140/epja/i2019-12688-3},
  url = {https://doi.org/10.1140/epja/i2019-12688-3}
}

@article{Nasirov2024,
  title = {Optimal colliding energy for the synthesis of a superheavy element with $Z=119$},
  author = {Nasirov, Avazbek and Kayumov, Bakhodir},
  journal = {Phys. Rev. C},
  volume = {109},
  issue = {2},
  pages = {024613},
  numpages = {10},
  year = {2024},
  month = {Feb},
  publisher = {American Physical Society},
  doi = {10.1103/PhysRevC.109.024613},
  url = {https://link.aps.org/doi/10.1103/PhysRevC.109.024613}
}

@article{Kayumov2022,
    author = "Kayumov, B. M. and Ganiev, O. K. and Nasirov, A. K. and Yuldasheva, G. A.",
    title = "{Analysis of the fusion mechanism in the synthesis of superheavy element 119 via the
    Cr54+Am243 reaction}",
    eprint = "2111.02861",
    archivePrefix = "arXiv",
    primaryClass = "nucl-th",
    doi = "10.1103/PhysRevC.105.014618",
    journal = "Phys. Rev. C",
    volume = "105",
    number = "1",
    pages = "014618",
    year = "2022"
}

@article{Fazio2003,
author = {Fazio ,Giovanni and Giardina ,Giorgio and Lamberto ,Antonino and Ruggeri ,Roberto and Saccà
,Carmelo and Palamara ,Rocco and I. Muminov ,Akhtam and K. Nasirov ,Avazbek and T. Yakhshiev ,Ulughbek
and Hanappe ,Francis and Materna ,Thomas and Stuttgé ,Louise},
title = {Entrance Channel Effect on the Formation of Heavy and Superheavy Nuclei},
journal = {J. Phys. Soc. Jpn.},
volume = {72},
number = {10},
pages = {2509-2522},
year = {2003},
doi = {10.1143/JPSJ.72.2509},
URL = {https://doi.org/10.1143/JPSJ.72.2509}
}

@article{Fazio2005,
  title = {Strong influence of the entrance channel on the formation of compound nuclei
  $^{216,222}\mathrm{Th}{}^{*}$ and their evaporation residues},
  author = {Fazio, G. and Giardina, G. and Mandaglio, G. and Ruggeri, R. and Muminov, A. I. and
  Nasirov, A. K. and Oganessian, Yu. Ts. and Popeko, A. G. and Sagaidak, R. N. and Yeremin, A. V. and
  Hofmann, S. and Hanappe, F. and Stodel, C.},
  journal = {Phys. Rev. C},
  volume = {72},
  issue = {6},
  pages = {064614},
  numpages = {14},
  year = {2005},
  month = {Dec},
  publisher = {American Physical Society},
  doi = {10.1103/PhysRevC.72.064614},
  url = {https://link.aps.org/doi/10.1103/PhysRevC.72.064614}
}

@article{Wheeler1939,
  title = {The Mechanism of Nuclear Fission},
  author = {Bohr, Niels and Wheeler, John Archibald},
  journal = {Phys. Rev.},
  volume = {56},
  issue = {5},
  pages = {426--450},
  numpages = {0},
  year = {1939},
  month = {Sep},
  publisher = {American Physical Society},
  doi = {10.1103/PhysRev.56.426},
  url = {https://link.aps.org/doi/10.1103/PhysRev.56.426}
}

@article{Weisskopf1937,
  title = {Statistics and Nuclear Reactions},
  author = {Weisskopf, V.},
  journal = {Phys. Rev.},
  volume = {52},
  issue = {4},
  pages = {295--303},
  numpages = {0},
  year = {1937},
  month = {Aug},
  publisher = {American Physical Society},
  doi = {10.1103/PhysRev.52.295},
  url = {https://link.aps.org/doi/10.1103/PhysRev.52.295}
}

@article{Pomorska2006,
  title = {Shell energy and the level-density parameter of hot nuclei},
  author = {Nerlo-Pomorska, Bo\ifmmode\dot{z}\else \.{z}\fi{}ena and Pomorski, Krzysztof and Bartel,
  Johann},
  journal = {Phys. Rev. C},
  volume = {74},
  issue = {3},
  pages = {034327},
  numpages = {8},
  year = {2006},
  month = {Sep},
  publisher = {American Physical Society},
  doi = {10.1103/PhysRevC.74.034327},
  url = {https://link.aps.org/doi/10.1103/PhysRevC.74.034327}
}

@article{Ignatyuk75,
  title = {Phenomenological description of energy dependence of the level density parameter},
  author = {A. V. Ignatyuk and G. N. Smirenkin and A. S. Tishin},
  journal = {Sov. J. Nucl. Phys.},
  volume = {21},
  issue = {3},
  pages = {485--490},
  numpages = {5},
  year = {1975},
  month = {Sep},
  publisher = { },
  doi = { },
  url = {https://inis.iaea.org/search/search.aspx?orig_q=RN:6208426}
}

@article{Lublin,
  title = {Optimal shapes and fission barriers of nuclei within the liquid drop model},
  author = {Ivanyuk, F. A. and Pomorski, K.},
  journal = {Phys. Rev. C},
  volume = {79},
  issue = {5},
  pages = {054327},
  numpages = {11},
  year = {2009},
  month = {May},
  publisher = {American Physical Society},
  doi = {10.1103/PhysRevC.79.054327},
  url = {https://link.aps.org/doi/10.1103/PhysRevC.79.054327}
}

@book{Hasse,
    author = "Hasse, R. W. and Myers, W. D.",
    title = "{Geometrical Relationships of Macroscopic Nuclear Physics}",
    year = "1988",
    publisher = "Springer Berlin, Heidelberg",
    url       = "https://doi.org/10.1007/978-3-642-83017-4"
}

@article{Sierk1986,
  title = {Macroscopic model of rotating nuclei},
  author = {Sierk, Arnold J.},
  journal = {Phys. Rev. C},
  volume = {33},
  issue = {6},
  pages = {2039--2053},
  numpages = {0},
  year = {1986},
  month = {Jun},
  publisher = {American Physical Society},
  doi = {10.1103/PhysRevC.33.2039},
  url = {https://link.aps.org/doi/10.1103/PhysRevC.33.2039}
}

@article{Giardina2000,
  title = { Effect of the entrance channel on the synthesis of superheavy elements},
  author = {Giardina, G. and Hofmann, S. and  Muminov, A. I. and  Nasirov, A. K.},
  journal = {The European Physical Journal A},
  volume = {8},
  issue = {2},
  pages = {205--216},
  numpages = {12},
  year = {2000},
  month = {Jul},
  publisher = {Springer},
  doi = {10.1007/s10050-000-4509-7},
  url = {https://link.aps.org/doi/10.1103/PhysRevLett.103.132502}
}

@article{Kemble1935,
  title = {A Contribution to the Theory of the B. W. K. Method},
  author = {Kemble, Edwin C.},
  journal = {Phys. Rev.},
  volume = {48},
  issue = {6},
  pages = {549--561},
  numpages = {13},
  year = {1935},
  month = {Sep},
  publisher = {American Physical Society},
  doi = {10.1103/PhysRev.48.549},
  url = {https://link.aps.org/doi/10.1103/PhysRev.48.549}
}

@article{Migdal1983,
author = {A.B. Migdal},
title = {Theory of the Finite Fermi--Systems and Properties of Atomic Nuclei},
 journal = {book},
publisher = {Nauka, Moscow},
year = {1983},
pages = {639}
}

@article{Adamian2000,
title = {Isotopic dependence of fusion cross sections in reactions with heavy nuclei},
author = {Adamian, G. G. and Antonenko, N. V. and Scheid, W.},
journal = {Nuclear Physics A},
volume = {678},
number = {1},
pages = {24-38},
year = {2000},
issn = {0375-9474},
doi = {https://doi.org/10.1016/S0375-9474(00)00317-1},
url = {https://www.sciencedirect.com/science/article/pii/S0375947400003171}
}

@article{Graeger2010,
  title = {Experimental study of the $^{238}\mathrm{U}$($^{36}\mathrm{S}$,$3\ensuremath{-}5n$)$^{269\ensuremath{-}271}\mathrm{Hs}$ reaction leading to the observation of $^{270}\mathrm{Hs}$},
  author = {Graeger, R. and Ackermann, D. and Chelnokov, M. and Chepigin, V. and D\"ullmann, Ch. E. and Dvorak, J. and Even, J. and Gorshkov, A. and He\ss{}berger, F. P. and Hild, D. and H\"ubner, A. and J\"ager, E. and Khuyagbaatar, J. and Kindler, B. and Kratz, J. V. and Krier, J. and Kuznetsov, A. and Lommel, B. and Nishio, K. and Nitsche, H. and Omtvedt, J. P. and Petrushkin, O. and Rudolph, D. and Runke, J. and Samadani, F. and Sch\"adel, M. and Schausten, B. and T\"urler, A. and Yakushev, A. and Zhi, Q.},
  journal = {Phys. Rev. C},
  volume = {81},
  issue = {6},
  pages = {061601},
  numpages = {5},
  year = {2010},
  month = {Jun},
  publisher = {American Physical Society},
  doi = {10.1103/PhysRevC.81.061601},
  url = {https://link.aps.org/doi/10.1103/PhysRevC.81.061601}
}

@article{AUDI2003,
title = {The Ame2003 atomic mass evaluation: (II). Tables, graphs and references},
journal = {Nuclear Physics A},
volume = {729},
number = {1},
pages = {337-676},
year = {2003},
note = {The 2003 NUBASE and Atomic Mass Evaluations},
issn = {0375-9474},
doi = {https://doi.org/10.1016/j.nuclphysa.2003.11.003},
url = {https://www.sciencedirect.com/science/article/pii/S0375947403018098},
author = {G. Audi and A. H. Wapstra and C. Thibault}
}

@article{Moller1995,
title = {Nuclear Ground-State Masses and Deformations},
author = {P. Moller and J.R. Nix and W.D. Myers and W.J. Swiatecki},
journal = {At. Data Nucl. Data Tables},
volume = {59},
number = {2},
pages = {185-381},
year = {1995},
issn = {0092-640X},
doi = {https://doi.org/10.1006/adnd.1995.1002},
url = {https://www.sciencedirect.com/science/article/pii/S0092640X85710029}
}

@article{Giardina2018,
title = {Uncertainties and understanding of experimental and theoretical results regarding reactions
forming heavy and superheavy nuclei},
author = {Giardina, G. and Mandaglio, G. and Nasirov, A. K. and A. Anastasi and F. Curciarello and G.
Fazio},
journal = {Nucl. Phys. A},
volume = {970},
pages = {169-207},
year = {2018},
issn = {0375-9474},
doi = {https://doi.org/10.1016/j.nuclphysa.2017.11.010},
url = {https://www.sciencedirect.com/science/article/pii/S037594741730475X},
}

@article{Kowal2010,
  title = {Fission barriers for even-even superheavy nuclei},
  author = {Kowal, M. and Jachimowicz, P. and Sobiczewski, A.},
  journal = {Phys. Rev. C},
  volume = {82},
  issue = {1},
  pages = {014303},
  numpages = {10},
  year = {2010},
  month = {Jul},
  publisher = {American Physical Society},
  doi = {10.1103/PhysRevC.82.014303},
  url = {https://link.aps.org/doi/10.1103/PhysRevC.82.014303}
}

@article{Jachimowicz2021,
title = {Properties of heaviest nuclei with 98≤Z≤126 and 134≤N≤192},
journal = {Atomic Data and Nuclear Data Tables},
volume = {138},
pages = {101393},
year = {2021},
issn = {0092-640X},
doi = {https://doi.org/10.1016/j.adt.2020.101393},
url = {https://www.sciencedirect.com/science/article/pii/S0092640X20300504},
author = {P. Jachimowicz and M. Kowal and J. Skalski}
}

@article{Mandaglio2018,
title = {Role of charged particle emission on the evaporation
residue formation in the 82Se+138Ba reaction leading to
the 220Th compound nucleus},
author = {Mandaglio, G. and Nasirov, A. K. and Anastasi, A. and Curciarello, F.  and 
Fazio, G. and Giardina, G.},
journal = {Nucl. Phys. A},
volume = {979},
pages = {204-214},
year = {2018},
issn = {0375-9474},
publisher = { Elsevier},
url ={https://doi.org/10.1016/j.nuclphysa.2018.09.057},
doi = {10.1016/j.nuclphysa.2018.09.057}
}

@article{Loveland2015,
title = { An experimentalist’s view of the uncertainties in understanding heavy element synthesis},
author = {Loveland, W.},
journal = {European Physical Journal A},
volume = {51},
pages = {120},
issue = {9},
year = {2015},
issn = {0375-9474},
doi = {10.1140/epja/i2015-15120-2},
url = {https://doi.org/10.1140/epja/i2015-15120-2},
}

@article{Jackson1956,
  title = {A schematic model for (p,xn) cross sections in heavy elements},
  author = {Jackson, J. D.},
  journal = {Can. J. Phys.},
  volume = {34},
  number = {8},
  pages = {767--779},
  year = {1956},
  doi = {10.1139/p56-087}
}

@book{Vandenbosch1973,
  author = {Vandenbosch, R. and Huizenga, J. R.},
  title = {{Nuclear Fission}},
  year = {1973},
  publisher = {Academic Press, New York}
}

@article{Moller2016,
  title = {Nuclear ground-state masses and deformations: FRDM(2012)},
  author = {M{\"o}ller, P. and Sierk, A. J. and Ichikawa, T. and Sagawa, H.},
  journal = {At. Data Nucl. Data Tables},
  volume = {109--110},
  pages = {1--204},
  year = {2016},
  doi = {10.1016/j.adt.2015.10.002}
}

@article{Moller2015,
  title = {Fission barriers at the end of the chart of the nuclides},
  author = {M{\"o}ller, P. and Sierk, A. J. and Ichikawa, T. and Iwamoto, A. and Mumpower, M.},
  journal = {Phys. Rev. C},
  volume = {91},
  issue = {2},
  pages = {024310},
  year = {2015},
  doi = {10.1103/PhysRevC.91.024310}
}

@article{Oganessian2004a,
  title = {Measurements of cross sections and decay properties of the isotopes of
           elements 112, 114, and 116 produced in the fusion reactions
           $^{233,238}$U, $^{242}$Pu, and $^{248}$Cm$+^{48}$Ca},
  author = {Oganessian, Yu. Ts. and others},
  journal = {Phys. Rev. C},
  volume = {70},
  issue = {6},
  pages = {064609},
  year = {2004},
  doi = {10.1103/PhysRevC.70.064609}
}

@article{Oganessian2004b,
  title = {Measurements of cross sections for the fusion-evaporation reactions
           $^{244}$Pu($^{48}$Ca,$xn$)$^{292-x}$114 and
           $^{245}$Cm($^{48}$Ca,$xn$)$^{293-x}$116},
  author = {Oganessian, Yu. Ts. and others},
  journal = {Phys. Rev. C},
  volume = {69},
  issue = {5},
  pages = {054607},
  year = {2004},
  doi = {10.1103/PhysRevC.69.054607}
}

@article{Gates2011,
  title = {First superheavy element experiments at the GSI recoil separator TASCA:
           The production and decay of element 114 in the $^{244}$Pu($^{48}$Ca,3-4n)
           reaction},
  author = {Gates, J. M. and others},
  journal = {Phys. Rev. C},
  volume = {83},
  issue = {5},
  pages = {054618},
  year = {2011},
  doi = {10.1103/PhysRevC.83.054618}
}

@article{SamarkRoth2023,
  title = {Spectroscopy of $^{288}$Fl and its decay products},
  author = {S{\aa}mark-Roth, A. and others},
  journal = {Phys. Rev. C},
  volume = {107},
  issue = {2},
  pages = {024301},
  year = {2023},
  doi = {10.1103/PhysRevC.107.024301}
}

@article{Gates2026,
  title = {Production and decay of flerovium isotopes in the
           $^{244}$Pu($^{48}$Ca,$xn$) reaction},
  author = {Gates, J. M. and others},
  journal = {Phys. Rev. C},
  volume = {113},
  issue = {2},
  pages = {024611},
  year = {2026},
  doi = {10.1103/PhysRevC.113.024611}
}

@article{Rahmatinejad2023,
  title = {Survival probabilities of superheavy nuclei},
  author = {Rahmatinejad, A. and Litvinova, E. and Bao, X. and Shneidman, T. M.},
  journal = {arXiv:2307.13437},
  year = {2023},
  eprint = {2307.13437},
  archivePrefix = {arXiv},
  primaryClass = {nucl-th}
}

@article{Kayumov2025,
  title = {Investigation of the formation of superheavy elements with atomic numbers 116 and 120 through $^{50}\mathrm{Ti}$-induced reactions},
  author = {Kayumov, Bakhodir},
  journal = {Phys. Rev. C},
  volume = {111},
  issue = {5},
  pages = {054619},
  numpages = {10},
  year = {2025},
  month = {May},
  publisher = {American Physical Society},
  doi = {10.1103/PhysRevC.111.054619},
  url = {https://link.aps.org/doi/10.1103/PhysRevC.111.054619}
}

@article{Volkov1978,
  title = {Deep inelastic transfer reactions --- the new type of reactions between complex nuclei},
  author = {Volkov, V. V.},
  journal = {Phys. Rep.},
  volume = {44},
  number = {2},
  pages = {93--157},
  year = {1978},
  doi = {10.1016/0370-1573(78)90200-4}
}

@article{Kawai2026,
  title = {Theoretical estimates for the synthesis of $Z=119$ superheavy nuclei
           with Ca, Ti, V, and Cr projectiles: Effects of reaction $Q$ values
           and mass-model dependence},
  author = {Kawai, K. and Aritomo, Y. and Nakajima, K. and Takagi, S.
            and Nishimura, N.},
  journal = {Phys. Rev. C},
  volume = {114},
  issue = {1},
  pages = {014612},
  year = {2026},
  doi = {10.1103/pjw1-1c5k},
}

@article{Kramers1940,
  title = {Brownian motion in a field of force and the diffusion model of
           chemical reactions},
  author = {Kramers, H. A.},
  journal = {Physica},
  volume = {7},
  pages = {284},
  year = {1940},
}

@article{Strutinsky1973,
  title = {The fission width of excited nuclei},
  author = {Strutinsky, V. M.},
  journal = {Phys. Lett. B},
  volume = {47},
  pages = {121},
  year = {1973},
}

@article{Oganessian2013,
  title = {Investigation of the ${}^{243}\mathrm{Am}+{}^{48}\mathrm{Ca}$ reaction products previously
  observed in the experiments on elements 113, 115, and 117},
  author = {Oganessian, Yu. Ts. and Abdullin, F. Sh. and Dmitriev, S. N. and Gostic, J. M. and
  Hamilton, J. H. and Henderson, R. A. and Itkis, M. G. and Moody, K. J. and Polyakov, A. N. and
  Ramayya, A. V. and Roberto, J. B. and Rykaczewski, K. P. and Sagaidak, R. N. and Shaughnessy, D. A.
  and Shirokovsky, I. V. and Stoyer, M. A. and Stoyer, N. J. and Subbotin, V. G. and Sukhov, A. M. and
  Tsyganov, Yu. S. and Utyonkov, V. K. and Voinov, A. A. and Vostokin, G. K.},
  journal = {Phys. Rev. C},
  volume = {87},
  issue = {1},
  pages = {014302},
  numpages = {10},
  year = {2013},
  month = {Jan},
  publisher = {American Physical Society},
  doi = {10.1103/PhysRevC.87.014302},
  url = {https://link.aps.org/doi/10.1103/PhysRevC.87.014302}
}

@article{Khuyagbaatar2020,
  title = {Search for elements 119 and 120},
  author = {Khuyagbaatar, J. and Yakushev, A. and D\"ullmann, Ch. E. and Ackermann, D. and Andersson,
  L.-L. and Asai, M. and Block, M. and Boll, R. A. and Brand, H. and Cox, D. M. and Dasgupta, M. and
  Derkx, X. and Di Nitto, A. and Eberhardt, K. and Even, J. and Evers, M. and Fahlander, C. and
  Forsberg, U. and Gates, J. M. and Gharibyan, N. and Golubev, P. and Gregorich, K. E. and Hamilton, J.
  H. and Hartmann, W. and Herzberg, R.-D. and He\ss{}berger, F. P. and Hinde, D. J. and Hoffmann, J.
  and Hollinger, R. and H\"ubner, A. and J\"ager, E. and Kindler, B. and Kratz, J. V. and Krier, J. and
  Kurz, N. and Laatiaoui, M. and Lahiri, S. and Lang, R. and Lommel, B. and Maiti, M. and Miernik, K.
  and Minami, S. and Mistry, A. K. and Mokry, C. and Nitsche, H. and Omtvedt, J. P. and Pang, G. K. and
  Papadakis, P. and Renisch, D. and Roberto, J. B. and Rudolph, D. and Runke, J. and Rykaczewski, K. P.
  and Sarmiento, L. G. and Sch\"adel, M. and Schausten, B. and Semchenkov, A. and Shaughnessy, D. A.
  and Steinegger, P. and Steiner, J. and Tereshatov, E. E. and Th\"orle-Pospiech, P. and Tinschert, K.
  and Torres De Heidenreich, T. and Trautmann, N. and T\"urler, A. and Uusitalo, J. and Wegrzecki, M.
  and Wiehl, N. and Van Cleve, S. M. and Yakusheva, V.},
  journal = {Phys. Rev. C},
  volume = {102},
  issue = {6},
  pages = {064602},
  numpages = {9},
  year = {2020},
  month = {Dec},
  publisher = {American Physical Society},
  doi = {10.1103/PhysRevC.102.064602},
  url = {https://link.aps.org/doi/10.1103/PhysRevC.102.064602}
}

@article{Hofmann2016,
  title = {Review of even element super-heavy nuclei and search for element 120},
  author = {Hofmann, S. and Heinz, S. and Mann, R. and Maurer, J. and Münzenberg, G. and Antalic, S.
  and Barth, W. and Burkhard, H. G. and Dahl, L. and Eberhardt, K. and Grzywacz, R. and Hamilton, J. H.
  and Henderson, R. A. and Kenneally, J. M. and Kindler, B. and Kojouharov, I. and Lang, R. and Lommel,
  B. and Miernik, K. and Miller, D. and Moody, K. J. and Morita, K. and Nishio, K. and Popeko, A. G.
  and Roberto, J. B. and Runke, J. and Rykaczewski, K. P. and Saro, S. and Scheidenberger, C. and
  Schött, H. J. and Shaughnessy, D. A. and Stoyer, M. A. and Thörle-Pospiech, P. and Tinschert, K. and
  Trautmann, N. and Uusitalo, J. and Yeremin, A. V.},
  journal = {Eur. Phys. J. A},
  volume = {52},
  issue = {6},
  pages = {180},
  numpages = {34},
  year = {2016},
  month = {June},
  publisher = {Springer Nature},
  doi = {10.1140/epja/i2016-16180-4},
  url = {https://doi.org/10.1140/epja/i2016-16180-4}
}

@article{Bao2015,
  title = {Isotopic dependence of superheavy nuclear production in hot fusion reactions},
  author = {Bao, X. J. and Gao, Y. and Li, J. Q. and Zhang, H. F.},
  journal = {Phys. Rev. C},
  volume = {92},
  issue = {3},
  pages = {034612},
  numpages = {7},
  year = {2015},
  month = {Sep},
  publisher = {American Physical Society},
  doi = {10.1103/PhysRevC.92.034612},
  url = {https://link.aps.org/doi/10.1103/PhysRevC.92.034612}
}

@article{Zhang2024,
  title = {Predictions of synthesizing elements with $Z=119$ and 120 in fusion reactions},
  author = {Zhang, Ming-Hao and Zhang, Yu-Hai and Zou, Ying and Wang, Chen and Zhu, Long and Zhang, Feng-Shou},
  journal = {Phys. Rev. C},
  volume = {109},
  issue = {1},
  pages = {014622},
  numpages = {14},
  year = {2024},
  month = {Jan},
  publisher = {American Physical Society},
  doi = {10.1103/PhysRevC.109.014622},
  url = {https://link.aps.org/doi/10.1103/PhysRevC.109.014622}
}

@article{Zhu2023,
  title = {Possibility to synthesize $Z=120$ superheavy nuclei with $Z>20$ projectiles},
  author = {Zhu, S. H. and Bao, Xiao Jun},
  journal = {Phys. Rev. C},
  volume = {108},
  issue = {1},
  pages = {014604},
  numpages = {7},
  year = {2023},
  month = {Jul},
  publisher = {American Physical Society},
  doi = {10.1103/PhysRevC.108.014604},
  url = {https://link.aps.org/doi/10.1103/PhysRevC.108.014604}
}

@article{Liu2013,
  title = {Possibility to produce element 120 in the ${}^{54}$Cr+${}^{248}$Cm hot fusion reaction},
  author = {Liu, Zu-Hua and Bao, Jing-Dong},
  journal = {Phys. Rev. C},
  volume = {87},
  issue = {3},
  pages = {034616},
  numpages = {6},
  year = {2013},
  month = {Mar},
  publisher = {American Physical Society},
  doi = {10.1103/PhysRevC.87.034616},
  url = {https://link.aps.org/doi/10.1103/PhysRevC.87.034616}
}

@article{Gates2024,
  title = {Toward the Discovery of New Elements: Production of Livermorium ($Z=116$) with $^{50}\mathrm{Ti}$},
  author = {Gates, J. M. and Orford, R. and Rudolph, D. and Appleton, C. and Barrios, B. M. and Benitez, J. Y. and Bordeau, M. and Botha, W. and Campbell, C. M. and Chadderton, J. and Chemey, A. T. and Clark, R. M. and Crawford, H. L. and Despotopulos, J. D. and Dorvaux, O. and Esker, N. E. and Fallon, P. and Folden, C. M. and Gall, B. J. P. and Garcia, F. H. and Golubev, P. and Gooding, J. A. and Grebo, M. and Gregorich, K. E. and Guerrero, M. and Henderson, R. A. and Herzberg, R.-D. and Hrabar, Y. and King, T. T. and Kireeff Covo, M. and Kirkland, A. S. and Kr\"ucken, R. and Leistenschneider, E. and Lykiardopoulou, E. M. and McCarthy, M. and Mildon, J. A. and M\"uller-Gatermann, C. and Phair, L. and Pore, J. L. and Rice, E. and Rykaczewski, K. P. and Sammis, B. N. and Sarmiento, L. G. and Seweryniak, D. and Sharp, D. K. and Sinjari, A. and Steinegger, P. and Stoyer, M. A. and Szornel, J. M. and Thomas, K. and Todd, D. S. and Vo, P. and Watson, V. and Wooddy, P. T.},
  journal = {Phys. Rev. Lett.},
  volume = {133},
  issue = {17},
  pages = {172502},
  numpages = {8},
  year = {2024},
  month = {Oct},
  publisher = {American Physical Society},
  doi = {10.1103/PhysRevLett.133.172502},
  url = {https://link.aps.org/doi/10.1103/PhysRevLett.133.172502}
}

@article{Li2022120,
  title = {Predictions for the synthesis of the $Z=120$ superheavy element},
  author = {Li, Jia-Xing and Zhang, Hong-Fei},
  journal = {Phys. Rev. C},
  volume = {106},
  issue = {3},
  pages = {034613},
  numpages = {8},
  year = {2022},
  month = {Sep},
  publisher = {American Physical Society},
  doi = {10.1103/PhysRevC.106.034613},
  url = {https://link.aps.org/doi/10.1103/PhysRevC.106.034613}
}

@article{Zagrebaev2008,
  title = {Synthesis of superheavy nuclei: A search for new production reactions},
  author = {Zagrebaev, Valery and Greiner, Walter},
  journal = {Phys. Rev. C},
  volume = {78},
  issue = {3},
  pages = {034610},
  numpages = {11},
  year = {2008},
  month = {Sep},
  publisher = {American Physical Society},
  doi = {10.1103/PhysRevC.78.034610},
  url = {https://link.aps.org/doi/10.1103/PhysRevC.78.034610}
}

@article{Tanaka2022,
author = {Tanaka ,Masaomi and Brionnet ,Pierre and Du ,Miting and Ezold ,Julie and Felker ,Kevin and
Gall ,Beno\^{\i}t J.P. and Go ,Shintaro and Grzywacz ,Robert K. and Haba ,Hiromitsu and Hagino ,Kouichi
and Hogle ,Susan and Ishizawa ,Satoshi and Kaji ,Daiya and Kimura ,Sota and King ,Thomas T. and Komori
,Yukiko and Lemon ,Raiden K. and Leonard ,Milan G. and Morimoto ,Kouji and Morita ,Kosuke and Nagae
,Daisuke and Naito ,Natsuki and Niwase ,Toshitaka and Rasco ,Bertis C. and Roberto ,James B. and
Rykaczewski ,Krzysztof P. and Sakaguchi ,Satoshi and Sakai ,Hideyuki and Shigekawa ,Yudai and Stracener
,Daniel W. and VanCleve ,Shelley and Wang ,Yang and Washiyama ,Kouhei and Yokokita ,Takuya},
title = {Probing Optimal Reaction Energy for Synthesis of Element 119 from $^{51}$V$+^{248}$Cm Reaction
with Quasielastic Barrier Distribution Measurement},
journal = {Journal of the Physical Society of Japan},
volume = {91},
number = {8},
pages = {084201},
year = {2022},
doi = {10.7566/JPSJ.91.084201},
URL = {https://doi.org/10.7566/JPSJ.91.084201},
eprint = {https://doi.org/10.7566/JPSJ.91.084201}
}

@article{Oganessian2023,
  title = {New isotope $^{276}\mathrm{Ds}$ and its decay products $^{272}\mathrm{Hs}$ and $^{268}\mathrm{Sg}$ from the $^{232}\mathrm{Th}+^{48}\mathrm{Ca}$ reaction},
  author = {Oganessian, Yu. Ts. and Utyonkov, V. K. and Shumeiko, M. V. and Abdullin, F. Sh. and Dmitriev, S. N. and Ibadullayev, D. and Itkis, M. G. and Kovrizhnykh, N. D. and Kuznetsov, D. A. and Petrushkin, O. V. and Podshibiakin, A. V. and Polyakov, A. N. and Popeko, A. G. and Rogov, I. S. and Sagaidak, R. N. and Schlattauer, L. and Shubin, V. D. and Solovyev, D. I. and Tsyganov, Yu. S. and Voinov, A. A. and Subbotin, V. G. and Bublikova, N. S. and Voronyuk, M. G. and Sabelnikov, A. V. and Bodrov, A. Yu. and Gan, Z. G. and Zhang, Z. Y. and Huang, M. H. and Yang, H. B.},
  journal = {Phys. Rev. C},
  volume = {108},
  issue = {2},
  pages = {024611},
  numpages = {7},
  year = {2023},
  month = {Aug},
  publisher = {American Physical Society},
  doi = {10.1103/PhysRevC.108.024611},
  url = {https://link.aps.org/doi/10.1103/PhysRevC.108.024611}
}

@article{Nasirov2026NPA,
    author = "Nasirov, Avazbek and Khusanov, Elzod and Kayumov, Bakhodir and Ganiev, Orifjon and Yusupov, Azimboy and Safarov, Abrorbek",
    title = "{The effect of the entrance channel properties in formation of the reaction products in heavy ion collisions}",
    doi = "10.1016/j.nuclphysa.2025.123262",
    journal = "Nucl. Phys. A",
    volume = "1065",
    pages = "123262",
    year = "2026"
}

@article{Kozulin2014,
  title = {Fission and quasifission of composite systems with $Z=108$--$120$: Transition from heavy-ion reactions involving S and Ca to Ti and Ni ions},
  author = {Kozulin, E. M. and Knyazheva, G. N. and Itkis, I. M. and Itkis, M. G. and Bogachev, A. A. and Krupa, L. and Oganessian, Yu. Ts. and Zagrebaev, V. I. and Rusanov, A. Ya. and Hanappe, F. and Materna, T. and Rowley, N. and Stuttge, L. and Giardina, G. and Kliman, J.},
  journal = {Phys. Rev. C},
  volume = {90},
  issue = {5},
  pages = {054608},
  year = {2014},
  doi = {10.1103/PhysRevC.90.054608}
}

\end{document}